\documentclass{iopconfser}
\usepackage{amsmath, amsthm, amssymb}
\usepackage{mathtools}
\usepackage{cases} 
\usepackage{braket}
\usepackage{upgreek} 

\usepackage{thmtools}
\usepackage{thm-restate}

\usepackage{hyperref}
\usepackage{float}

\usepackage{graphicx}
\usepackage{adjustbox,lipsum}
\usepackage{xcolor}

\usepackage{tikz}
\usetikzlibrary{quantikz}
\usetikzlibrary{decorations.text}
\usetikzlibrary{decorations.pathreplacing}

\usepackage{pgfplots}
\usepgfplotslibrary{groupplots} 
\usepackage{anyfontsize} 

\usepackage{rotating}

\usepackage{algorithm}
\usepackage{algpseudocode}
\usepackage{algorithmicx}

\usepackage{pdflscape}

\newtheorem{remark}{Remark}
\newtheorem{lemma}{Lemma}
\newtheorem{definition}{Definition}

\newtheorem{theorem}{Theorem}
\newtheorem{problem}{Problem}

\definecolor{darkgray176}{RGB}{176,176,176}

\definecolor{c3}{rgb}{0.13, 0.26, 0.12}
\definecolor{c2}{rgb}{0.8, 0.47, 0.13}
\definecolor{c1}{rgb}{0.0, 0.0, 1.0}

\newcommand{\cheby}{\mathrm{T}}
\newcommand{\chebyt}{\mathrm{U}}

\newcommand{\dom}[1]{\mathcal{D}_{1}}
\newcommand{\cost}[1]{\mathcal{C}\left[#1\right]}
\newcommand{\linfnorm}[1]{\left|\left|#1\right|\right|_{\infty}}
\newcommand{\opnorm}[1]{\left|\left|#1\right|\right|_{op}}

\usepackage{hyperref}
\usepackage{cleveref}
\crefname{problem}{problem}{problem}

\usetikzlibrary{shapes.geometric, arrows}
\tikzstyle{io} = [trapezium,  minimum width=0.5cm, minimum height=0.5cm, text centered, fill=c1!5, draw=black, ]
\tikzstyle{process} = [rectangle, minimum width=3cm, minimum height=1cm, fill=c2!5,text centered, draw=black,]
\tikzstyle{decision} = [trapezium, minimum width=0.5cm, minimum height=0.5cm, text centered,  fill=c1!5, draw=black,]
\tikzstyle{arrow} = [thick,->,>=stealth]

\usepackage{caption}
\usepackage{subcaption}

\pgfplotsset{compat=1.18} 

\begin{document}
\title{The Hitchhiker's Guide to QSP pre-processing}
\author{S. E. Skelton$^{1}$}
\affil{$^{1}$Institute for Theoretical Physics, University of Leibniz Hannover}
\email{shawn-skelton@itp.uni-hannover.de}

\begin{abstract}
    Quantum signal processing relies on a historically costly pre-processing step, "QSP-processing/phase-factor finding." QSP-processing is now a developed topic within quantum algorithms literature, and a beginner accessible review of QSP-processing is overdue. This work provides a whirlwind tour through QSP conventions and pre-processing methods, beginning from a pedagogically accessible QSP convention. We then review QSP conventions associated with three common polynomial types: real polynomials with definite parity, sums of reciprocal/anti-reciprocal Chebyshev polynomials, and complex polynomials. We demonstrate how the conventions perform with respect to three criteria: circuit length, polynomial conditions, and pre-processing methods.\\
We then review the recently introduced Wilson method for QSP-processing and give conditions where it can succeed with bound error. Although the resulting bound is not computationally efficient, we demonstrate that the method succeeds with linear error propagation for relevant target polynomials and precision regimes, including the Jacobi-Anger expansion used in Hamiltonian simulation algorithms. We then apply our benchmarks to three QSP-processing methods for QSP circuits and show that a method introduced by Berntson and S{\"u}nderhauf outperforms both the Wilson method and the standard optimization strategy for complex polynomials. 
\end{abstract}

\tableofcontents
\pagenumbering{arabic}
\section{Introduction}
In their usual presentations, the most common quantum algorithms are conceptually disconnected. Practitioners learn and build intuition for individual routines, for example the quantum Fourier transform, state preparation schemes, linear solvers, and Hamiltonian simulation algorithms. Having a template within which to develop quantum algorithms would be a desirable alternative both for pedagogical and practical reasons; developments, modifications, and simplifications could then be designed within one unified description. In the past five years, two such proposed templates,  quantum signal processing (QSP) and the related Quantum singular value transformations (QSVT), have ascended into  popular models for developing quantum algorithms.\\

QSP and QSVT share some transparent advantages; all standard quantum algorithms can be built using one of them with similar asymptotic scaling compared to 'usual' algorithms for the same tasks \cite{martyn_grand_nodate, gilyen_quantum_2019, rall_faster_2021}. Furthermore, the precision of the algorithm relates to the circuit length in a very intuitive manner, and it is extremely simple to tailor the precision of the quantum algorithm with QSP. An increasing literature of QSP and QSVT applications, extensions, and modifications further testifies to these heuristics' increasing popularity.\\

 At its simplest,  QSP/QSVT is a template for computing some polynomial of the eigenvalues/singular values of an operator on a quantum circuit. The polynomials used in QSP/QSVT are often truncated forms of analytically bounded expansions, most commonly, trigonometric series. Alternately,  one can use the Remez algorithm to identify a suitable real polynomial \cite{dong_efficient_2021} approximating the desired function. To actually design an algorithm with QSP, the user must also define an oracle to which they want to apply this polynomial. Then, the QSP circuit iterates between controlled versions of the oracle and a series of rotation matrices on a control qubit, preparing a polynomial of the eigenvalues of the oracle on one branch of the evolved state.\\
 
 These rotation matrices fully specify the QSP circuit for a given polynomial; to find them, QSP and QSVT share a classical pre-processing step. This "QSP-processing"\footnote{A common term for this is `phase factor finding', which we have chosen to avoid because not all QSP-conventions have $SU(2)$ rotations defined by a single angle.} has expanded into its own sub-discipline of study \cite{chao2020finding, Haah2019product, dong_efficient_2021, Ying2022stablefactorization} across several popular QSP conventions \cite{Haah2019product, dong_efficient_2021, motlagh_generalized_2023}. 
The QSP-processing,  QSP convention choice, and `suitable'\footnote{`Suitable' is convention dependent, and in our case defined precisely in \cref{lemma:QSP_gilyen}\textemdash \cref{lemma:QSP_Motlagh}.} polynomial approximation all introduce specific conditions that must be jointly satisfied for the QSP/QSVT algorithm to succeed.\\

 However, the explosion of research in QSP means that the standard framework of QSP is often not the optimal choice. Currently, users cannot easily discern the computational advantages and disadvantages of a particular convention without a deep understanding of the literature. This paper offers three contributions aimed at making the process of getting into QSP easier for beginners. We provide a review of QSP-processing strategies; a detailed discussion of one QSP-processing strategy, including a full error analysis; and a comparison between three techniques for generalized QSP (G-QSP) circuits with reciprocity restrictions on the input polynomials. This specification encompasses the majority of QSP applications, and two of the three methods compared can be used in the fully general G-QSP setting.\\

We schematize QSP conventions into three types. The most common convention is angle-QSP \cite{gilyen_quantum_2019, martyn_grand_nodate}, with specialization `symmetric QSP'\cite{lin_lecture_2022}. This has the most restricted polynomial approximations; historically,  it also had the most successful QSP processing methods. In angle-QSP, linear combinations of unitaries (LCU) techniques \cite{berry_hamiltonian_2015, gilyen_quantum_2019}  can be used to construct arbitrary complex polynomials on the circuit. G-QSP is the convention that allows for the most general polynomials and has an optimal circuit length for Hamiltonian simulation \cite{berry2024doubling}. Between these two, Laurent-QSP\footnote{In this work, `Laurent-QSP' is always used to distinguish Laurent polynomials with reciprocity restrictions discussed below. Elsewhere, `Laurent-QSP' has been used to refer to a G-QSP circuit that builds any Laurent polynomial without imposing restrictions on the reciprocity. } allows polynomials for all important applications but has a less successful QSP-processing track record than symmetric QSP.\\

We compare the performance of three QSP-processing methods for Laurent-QSP, which is the original QSP convention and is general enough to implement most of the polynomials used in QSP algorithms. We compare a straightforward optimization strategy from \cite{motlagh_generalized_2023} (optimization), the algorithm from \cite{skelton2024harmlessmethodsqspprocessinglaurent} (Wilson), and a scheme introduced by Berntson and S{\"u}nderhauf  (BerS{\"u}n) which relies primarily on a series of Fast Fourier Transform steps \cite{berntson2024complementarypolynomialsquantumsignal}. We find that the BerS{\"u}n method is the consistent best-performing choice, with low error and fast completion times. We also prove that like the root-finding based method  \cite{Haah2019product}, the method in \cite{skelton2024harmlessmethodsqspprocessinglaurent} converges to a successful solution with arbitrary precision.  \\

Because the success of any QSP-processing strategy relies on the relevant polynomial approximation, target QSP convention, and acceptable error, we devote the majority of this paper to a high-level review of QSP techniques and conventions. In \cref{sec:applications} and \cref{sec:qsp_preliminaries}, we motivate QSP and define commonly used QSP-conventions, then \cref{sec:circuits} gives a high-level circuit description of QSP including common oracles and linear combination of unitaries strategies. This circuit-level analysis is crucial to comparing the costs of respective QSP conventions. Next, \cref{sec:qsp_processing} summarizes existing techniques for QSP-processing. The technical contributions of this work are \cref{sec:fejer}, which gives our bounded error analysis of Laurent-QSP, \cref{thrm:laurent_qsp_with_processing}, and \cref{sec:results}, which applies our results to polynomials of practical interest.

\subsection{QSP Applications and Modifications}\label{sec:applications}
The explosion of interest in QSP is now being followed by simulations and implementations aimed at probing its relative success against other quantum algorithms. At least one prominent recent work shied away from QSP \cite{costaoptimal2022} for concerns over its pre-processing success, and resource estimates comparing QSP to alternate techniques will likely be a crucial test of whether the template retains long-term popularity.\\

Because the QSP heuristic allows one to obtain an extremely short circuit at the cost of low precision, some QSP routines have been proposed for near-term benchmarking \cite{Dong_2022, dong_random_2021} or intermediate quantum algorithms \cite{dong_ground_2022}. There has been at least one simple experimental realization of QSP \cite{kikuchi_realization_2023}, as well as resource estimates for using QSP to simulate gauge theories \cite{kane2024blockencodingsignalprocessing, crane2024hybridoscillatorqubitquantumprocessors}. Additionally, the clear trade-off between the polynomial's degree and the circuit length makes QSVT a useful framework to evaluate depth\textendash precision trade-offs \cite{magano_simplifying_2022}. \\

Basic QSVT routines have been developed for phase, energy, and amplitude estimation problems \cite{rall_faster_2021, gilyen_quantum_2019, martyn_grand_nodate, steudtner_fault-tolerant_2023,magano_simplifying_2022}, state preparation \cite{laneve_robust_2023, mcardle_quantum_2022}{and transformation} \cite{guo2024nonlineartransformationcomplexamplitudes, rattew2023nonlinear}, and ground state problems \cite{dong_ground_2022}. QSVT is increasingly popular for applications of linear systems problems \cite{chakraborty_quantum_2022, jennings_efficient_2023, lin_optimal_2020, novikau_simulation_2022} and Hamiltonian simulation \cite{novikau_quantum_2022, martyn_efficient_2023, toyoizumi_hamiltonian_2023}. Some algorithms for {property measuring problems} use QSVT subroutines \cite{li_unified_2022, wang_new_2022, gilyen_improved_2022, quek_fast_2021}, as well as applications in quantum finance \cite{stamatopoulos_derivative_2023}, cryptography \cite{lombardi_post-quantum_2021}, constructing Petz recovery channels \cite{gilyen_quantum_2022}, and metrology \cite{dong2022heisenberglimitquantummetrology}. QSVT has also been studied in the context of "dequantized" classical algorithms \cite{gharibian_dequantizing_2022, bakshi_improved_2023}. \\

In recent years, an increasing number of results have attempted QSP-processing either "direct" methods \cite{toyoizumi_hamiltonian_2023, magano_simplifying_2022, novikau_quantum_2022} or optimization \cite{novikau_simulation_2022, stamatopoulos_derivative_2023, kikuchi_realization_2023, dong_random_2021}. However, historically these applications often have to adjust simulation parameters to avoid high degree polynomials \cite{novikau_quantum_2022,toyoizumi_hamiltonian_2023}, or restrict themselves to relatively low precision\textemdash in the best-identified case, $10^{-10}$ in \cite{toyoizumi_hamiltonian_2023} but more commonly, $\mathcal{O}(10^{-3})$ \cite{dong_random_2021, kikuchi_realization_2023} experimentally or $\mathcal{O}(10^{-6})$ \cite{novikau_quantum_2022, novikau_simulation_2022} in simulations. Recursive methods are sometimes used to obtain more complicated functional approximations \cite{gomes2024multivariableqspbosonicquantum}, \cite{Dong_2022}, and sometimes numerically approximating QSP parameters can be successful \cite{novikau2024estimatingqsvtanglesmatrix}. The processing methods discussed in this work are able to outperform these implementations by orders of magnitude, reflecting an extremely recent expansion of QSP-processing capacities for both real and complex target functions.\\

 QSP and QSVT have both been modified or reinterpreted, and a semantic embedding for QSVT has been proposed \cite{rossi_semantic_2023}, as well as a syntax for QSVT within functional programming \cite{rossi2023modular}. A few works have explored simplified settings to decrease circuit length \cite{borns-weil_quantum_2023, mizuta_recursive_2023}. Still others incorporate variable time amplitude estimation into QSVT algorithms \cite{li_unified_2022, chakraborty_quantum_2022, chakraborty_power_2019, lombardi_post-quantum_2021}. QSP modifications are still an important area of work, notably for multivariable functions \cite{mori2024commentmultivariablequantumsignal, rossi_multivariable_2022, németh2023variantsmultivariatequantumsignal,laneve2024multivariatepolynomialsachievablequantum}, and QSP-like procedures for more complicated Lie groups \cite{rossi2023modular, bastidas2023quantum}, \cite{bastidas2024complexificationquantumsignalprocessing}. \textit{Stochastic}-QSP \cite{martyn2024halvingcostquantumalgorithms} leverages randomized compiling to shorten the average query complexity, by averaging over an ensemble of polynomials which converges to an approximation of the desired function. \\

 There are also some proposed QSP generalizations, in particular, quantum eigenvalue processing from \cite{low2024quantumeigenvalueprocessing} and Fourier QSP from \cite{silva_fourier-based_2022}, which we could not discuss in detail. Fourier QSP uses oracle $e^{i\tau H}$, rather than  $e^{i\tau \arcsin(H)}$, to apply functions approximated by complex Fourier series. This approach is more general than any approach except G-QSP and has a root-finding QSP-processing strategy similar to the ones discussed herein.\\
 
 An alternate approach to QSP is using quantum eigenvalue processing, which applies linear combinations of Chebyshev polynomials to block-encodings of eigenvalues in subsets of the complex plane. In this way, it is morally similar to QSP. However, here, the strategy for applying the sum of Chebyshev polynomials is to encode each coefficient with a state preparation scheme and then implement a series of LCU steps. This approach also highlights one potential challenge to QSP\textemdash when quantum algorithms can be constructed with either LCU or QSP/QSVT techniques, it is an open question whether QSVT could unambiguously and consistently outperform LCU.

\section{QSP Preliminaries}\label{sec:qsp_preliminaries}
We review QSP\footnote{This section closely follows the introductory materials of companion paper \cite{skelton2024harmlessmethodsqspprocessinglaurent}} using the framework and notation introduced in \cite{low_hamiltonian_2019, Haah2019product}. This convention is the focus of our work and also provides the clearest physical intuition for QSP. 
Given  $\{\ket{p}\bra{p},\ket{q}\bra{q}\}$, a basis set in $SU(2)$, we consider a controlled operation on unitary $U$ \textit{controlled by an ancilla qubit with respect to the $p, q$ basis },
\begin{equation}
    C_{p}U=\ket{p}\bra{p}\otimes U+\ket{q}\bra{q}\otimes I=VC_0UV^{\dag}.
\end{equation}
 Above, $C_0U$ is the $0$-controlled $U$ gate for arbitrary unitary $U$ and $V\in SU(2)$ is the unitary transformation between the computational basis $\{\ket{p},\bra{q}\}$, that is, $V\ket{0}=\ket{p}$, $V\ket{1}=\ket{q}$. 
We denote the eigenvectors, eigenvalues of $U$, as $\ket{\theta}$, $e^{i\theta}$ respectively. Acting $C_pU$ on some initial state $\ket{\psi_0}=\ket{0}\otimes \sum c_k\ket{\theta_k}$, where $\sum c_k^2=1$ are a set of up to $\text{rank}(U)$ coefficients, we see that
\begin{align}
    C_pU\ket{\psi_0}&=\sum_kc_k\left(e^{i\theta}\ket{p}\bra{p}+\ket{q}\bra{q}\right)\ket{0}\ket{\theta_k}\\
    &=\sum_kc_ke^{i\theta/2}\left(e^{i\theta/2}\ket{p}\bra{p}+ e^{-i\theta/2}\ket{q}\bra{q}\otimes U\right)\ket{0}\ket{\theta_k}.
\end{align}
Assuming we also have access to the inverse operation $C_pU^{\dag}$, then global phases $e^{i\theta/2}, e^{-i\theta/2}$ cancel after consecutive $C_{p}U, C_{p'}U^{\dag}$ operations, resulting in 
\begin{multline}
    \sum_kc_k\left(e^{i\theta/2}\ket{p}\bra{p}+e^{-i\theta/2}\ket{q}\bra{q}\right) \left(e^{-i\theta/2}\ket{p'}\bra{p'}+e^{i\theta/2}\ket{q'}\bra{q'}\right)\ket{\theta_k}.
\end{multline} 
If the input state is an eigenvector of $U$, then the action of a sequence of $C_{p_j}UC_{p_{j+1}}U^{\dag}$ operations on the circuit is equivalent to the following sequence on the ancillary 
\begin{align}
    F(e^{i\theta/2})&=E_0E_{p_1}(e^{i\theta/2})E_{p_2}(e^{i\theta/2})...E_{p_{2n}}(e^{i\theta/2}),\\
    E_p(e^{i\theta/2})&=e^{i\theta/2}\ket{p}\bra{p}+e^{-i\theta/2}\ket{q}\bra{q}.
\end{align}
$F(e^{i\theta/2})\in SU(2)$ is a polynomial (with coefficients determined by products of $\braket{p_k|p_{j}}, \braket{p_{k}|q_{j}},...$) with Pauli decomposition
\begin{equation}
  F(e^{i\theta})=\mathcal{A}(e^{i\theta})\mathcal{I}+i\mathcal{B}(e^{i\theta})\sigma_X+i\mathcal{C}(e^{i\theta})\sigma_Y+i\mathcal{D}(e^{i\theta})\sigma_Z.
\end{equation}
Laurent polynomials $\mathcal{A, B, C, D}$ are fully determined by the set of $SU(2)$ projectors $\cup_{k=0}^{2n}\{\ket{p_k}\bra{p_k}\}$, are degree at most $n$, are real and have definite reciprocity on the unit circle, and satisfy 
\begin{equation}\label{eq:Haah_sqr_sum_constraint}
    \mathcal{A}^2(e^{i\theta})+\mathcal{B}^2(e^{i\theta})+\mathcal{C}^2(e^{i\theta})+\mathcal{D}^2(e^{i\theta})=1
\end{equation}
for all $\theta\in \mathbb{R}$. Once we measure the ancillary, then some combination of  $\mathcal{A, B, C, D}$ will be prepared on the rest of the circuit. The extension to $\ket{\psi_0}\neq \ket{\theta_k}$ is entirely straightforward. For example, if we project into the $\ket{+}\bra{+}$ subspace of the ancilla qubit, then the remainder of the circuit will be prepared in
\begin{equation}
    \ket{\psi_{QSP}}=\sum_kc_k\left(\mathcal{A}(e^{i\theta})+i\mathcal{B}(e^{i\theta})\right)\ket{\theta_k}.
\end{equation}
We can immediately define the QSP operator, 
\begin{align}\label{eq:QET_operator}
U_{QSP}&=E_0\otimes I_{\text{dim}(U)}\prod_{k=1}^{n} C_{p_k}UC_{p_{k+1}}U^{\dag},
\end{align}
whose circuit is given in \cref{fig:QET_basic_circuit}. The desired polynomial transformation is
\begin{align}
\sum_{\theta}\mathcal{P}(\theta)\ket{\theta}\bra{\theta}=\bra{\cdot}E_0\prod_{k=1}^{n} C_{p_k}UC_{p_{k+1}}U^{\dag}\ket{\cdot}
\end{align}
where $\bra{\cdot}\cdot \ket{\cdot}$ denotes post-selection on some basis measurement on the QSP ancilla qubit\footnote{Concretely: if we want to measure observable $O$ after preparing $\ket{\psi_{QSP}}=\sum_{\theta}\mathcal{P}(\theta)\ket{\theta}\bra{\theta}\ket{\psi_{initial}}$, then we first have to measure the QSP ancilla in the Pauli $\sigma_X$ basis, and if we obtain outcome $\ket{+}$ we can proceed to measure the system qubits. If we obtain outcome $\ket{-}$, then we have prepared the system qubits in $\mathcal{A}(z)-i\mathcal{B}(z)$ instead of the desired outcome.}. The \textit{target polynomial} $\mathcal{P}$ is a (complex) linear combination of $\{\mathcal{A, B, C, D}\}$ formed by a measurement on the ancillary qubit, usually the Hadamard basis state $\bra{+}\cdot \ket{+}$. This produces $\mathcal{P}(z)=\mathcal{A}(z)+i\mathcal{B}(z)$ and we often stipulate $\mathcal{A}$ is reciprocal and $\mathcal{B}$ is anti-reciprocal. See \cref{sec:poly_approx} for a review of Laurent polynomials.\\

So far, we have shown that a series of controlled operations and transformations $\{V_p\}$ prepares a complex Laurent polynomial of the eigenvalues of $U$. With a complete set of polynomials $\{\mathcal{A, B, C, D}\}$, constructing $\{\ket{p_j}\bra{p_j}, \ket{q_j}\bra{q_j}\}$ is straightforward. Define the matrix-valued  polynomial $F^{2n}(e^{i\theta/2})=\sum_{k=-2n}^{2n} C_{k}^{2n}e^{ik\theta/2}$, with matrix-valued coefficient list $\{C_{k}^{2n}\}$. The superscripts denote the highest degree with respect to $e^{i\theta/2}$. Then, we recursively define 
\begin{align}
     \ket{p_j}\bra{p_j}&=\frac{\left(C^{j}_{-n}\right)^{\dag}C^{j}_{-n}}{\text{Tr}\left(\left(C_{-n}\right)^{\dag}C_{-n}\right)}\\
     &\ket{q_j}\bra{q_j}=I-\ket{p_j}\bra{p_j}\\
     F^{j-1}(e^{i\theta/2})&=F^{j}(e^{i\theta/2})E_j(e^{i\theta/2}).
\end{align}
Lemma \ref{lemma:QSP_Haah} summarizes this into the Laurent-QSP convention.\\

A central difficulty of QSP in practice is deriving complimentary polynomials. Usually, QSP is constructed beginning from target polynomial $\mathcal{P}(\theta)=\mathcal{A}(\theta)+i\mathcal{B}(\theta)$, and defining complementary polynomial  $i\sqrt{1-\cos^2\theta}\mathcal{Q}(\theta)=-\mathcal{C}(\theta)+i\mathcal{D}(\theta)$. To ensure $\mathcal{C, D}$ fit the requirement \cref{eq:Haah_sqr_sum_constraint}, we have to solve the following problem. 
\begin{problem}[QSP-completion Problem]\label{prob:qsp_poly_constraint}
    Beginning with some polynomial $P$, given as a coefficient list, we want to identify some $Q$ such 
    \begin{equation}
        \left|Q(\cdot)\right|^2=1-\left|P(\cdot)\right|^2
    \end{equation}
    Here  $P(\cdot)$ denotes a polynomial in $z, x,$ or $\theta$
\end{problem}
We have phrased \cref{prob:qsp_poly_constraint} generically to stress that \textit{all} QSP conventions have such a step, and the associated polynomial properties restrict available solution methods. QSP conventions have historically managed a trade-off between the restrictions on allowed polynomials $\mathcal{P}(\cdot)$ and the conceptual and computational ease of identifying a complementary polynomial $\mathcal{Q}(\cdot)$ in \cref{prob:qsp_poly_constraint}. Then, the error in approximating $\mathcal{Q}$ propagates through the decomposition step. Thus, in practice, we cannot obtain $\{E\}$ exactly and instead want to identify a suitable set of $\{E\}$ which is some $\epsilon_{qsp}$-close to $\mathcal{P}(e^{i\theta})$.

\subsection{QSP Polynomials}
We assume the most general polynomial to implement in QSP has the form $\mathcal{P}: U(1)\rightarrow \mathbb{C}$, such that $|\mathcal{P}(z)|\leq 1$. Using the standard conversion valid on $U(1)$, $z=e^{i\theta}\leftrightarrow x=\cos\theta$,
 \begin{align}
     {P}(z)&=\sum_{k=0}^n a_kz^k=\sum_k^n a_k\left(x+i\sqrt{1-x^2}\right)^k\nonumber\\
     &=\sum_k^n\Tilde{a}_kx^k={P}'(x)
 \end{align}
 for coefficients $a_k, \Tilde{a}_k\in \mathbb{C}$.  Interested readers can refer to \cite{tang_cs_2023} for an overview of Chebyshev combinations for polynomial approximations in QSVT.
 Decomposed into even/odd real-on-circle polynomials,
 \begin{align}\label{eq:poly4termdecomp}
     {P}'(x)&=f_{R,E}(x)+f_{R, O}(x)+if_{I,E}(x)+if_{I, O}(x)
 \end{align}
Note that $|{P}'|=\sum_i|f_i|^2\leq 1\Rightarrow |f_i|\leq 1$, for $i\in\{(R, E), (R, O), (I, E), (I, O)\}$.\\

Historically, the strongest methods for solving QSP-processing applied only to $f_{real}:[-1, 1]\rightarrow \mathbb{R}$ \cite{dong_efficient_2021, gilyen_quantum_2019}. In this case, QSP-processing finds a complex function $P_c(x)=f_{real}(x)+if_{imag}(x)$ and a $Q(x)$ (which can be assumed real \cite{dong_efficient_2021}) and then obtains $f_{real}=\frac{1}{2}(P_c(x)+P_c^{*}(x))$ by adding two QSP circuits using a linear combination of unitaries technique. This provides a reasonable strategy to implement a generic complex polynomial $P(x)\rightarrow \mathbb{C}$. One can solve QSP-processing for $f_{R,E}, f_{I,E}, f_{R,O}, f_{I,O}$, and then use an LCU to add the approximations on a circuit. 

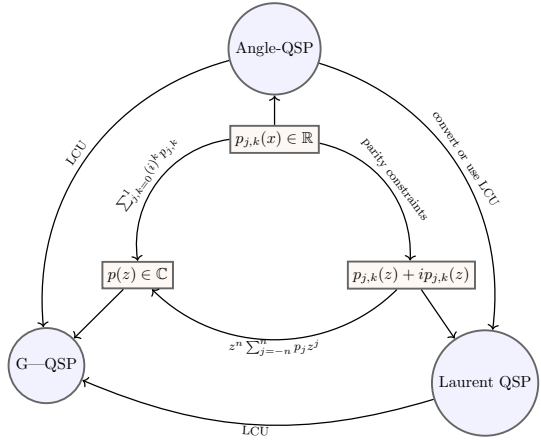
\begin{figure}[h]
    \centering
    \scalebox{0.6}{
      \begin{tikzpicture}[
roundnode/.style={circle, draw=black!60, fill=c1!5, very thick, minimum size=7mm},
squarednode/.style={rectangle, draw=black!60, fill=c2!5, very thick, minimum size=5mm}, 
]
    \node[squarednode] (one) at (0,3)   {$p_{j,k}(x)\in \mathbb{R}$}; 
    \node[squarednode] (two) at  (3,0) {$p_{j,k}(z)+ip_{j,k}(z)$};
    \node[squarednode] (three) at (-3,0) {$p(z) \in \mathbb{C}$};
    \draw [->,thick,postaction={decorate,decoration={raise=1ex,text along path,text align=center, text={|\footnotesize|{parity constraints}{}}}}] (one) to [bend left=40] (two);
    \draw [<-,thick,postaction={decorate,decoration={raise=1.5ex,text along path,text align=center,text={|\footnotesize|{$\sum_{j, k=0}^1 (i)^{k}p_{j, k}$}{}}}}] (three) to [bend left=45] (one);
    \draw [<-,thick,postaction={decorate,decoration={raise=-1.5ex,text along path,text align=center,text={|\footnotesize|{$z^n\sum_{j=-n}^n p_jz^j$}{}}}}] (three) to [bend right=45] (two);
    \node[roundnode] (four) at (-5,-2)   {G—QSP}; 
    \node[roundnode]  (five) at (5,-2, )  {Laurent QSP};
    \node[roundnode]  (six) at (0, 5)  {Angle-QSP};
    \draw [<-,thick,postaction={decorate,decoration={raise=-1.5ex,text along path,text align=center,text={|\footnotesize|{LCU}{}}}}] (four) to [bend left=-20] (five);
    \draw [->,thick,postaction={decorate,decoration={raise=1ex,text along path,text align=center,text={|\footnotesize|{convert or use LCU}{}}}}] (six) to [bend left=40] (five);
    \draw [<-,thick,postaction={decorate,decoration={raise=1ex,text along path,text align=center,text={|\footnotesize|{LCU}{}}}}] (four) to [bend left=40] (six);
    \draw [->,thick,] (one) to (six);
    \draw [->,thick] (two) to  (five);
    \draw [->,thick,] (three) to (four);
    \end{tikzpicture}  
    }
    \caption{reproduced from \cite{skelton2024harmlessmethodsqspprocessinglaurent}. Commonly used polynomials for QSP and associated QSP-conventions, assuming all polynomials obey $\linfnorm{p}\leq 1$. While some complex polynomials can be solved with Angle-QSP, the convention is predominantly used for real ($k=0$) or purely imaginary ($k=1$) polynomials with definite parity ($j=0, 1$). Coordinate transformation $x=\cos\theta\rightarrow z^{i\theta}$ allows polynomials of the form $p_{0, j}(z)+p_{1, j}(z)$ to be solved with Laurent-QSP. Complex polynomials (or Laurent polynomials) can be constructed as combinations of Laurent or real polynomials fitting the parity restrictions above. In some cases, Laurent and Angle QSP circuits are directly convertible. Linear combination of unitaries (LCU) techniques can be used on Angle or Laurent-QSP circuits to obtain more general polynomials. }
    \label{fig:convention_sketch}
\end{figure}

\subsection{QSP Conventions}
We distinguish three QSP conventions (`angle-QSP'/`symmetric-QSP,' `Laurent-QSP,' and `G-QSP') and the angle-finding methods available for each. These represent the most common approaches (and corresponding restrictions) for QSP. 

\subsubsection{Angle-QSP}
There is a family of QSP conventions we collectively call angle-QSP, such that all $V_{j}^{\dag}V_{j+1}$ pairs are rotations about one Pauli gate and thus specified by a single angle $\phi_{j}$. An angle-QSP circuit is fully specified by an angle set $\Phi_{n+1}=\{\phi_{j}\}$. When the desired polynomial approximation is real, symmetric-QSP \cite{dong_efficient_2021}  and related 'infinite-QSP' \cite{dong_infinite_2022} are specializations that can improve QSP-processing.\\

\begin{lemma}[Angle QSP, para from \cite{gilyen_quantum_2019}]\label{lemma:QSP_gilyen}
    Define 
    \begin{equation}
        W(x)\:=\begin{bmatrix}
        x & i\sqrt{1-x^2}\\
        i\sqrt{1-x^2} & x
    \end{bmatrix}=e^{i\arccos(x)\sigma_X},
    \end{equation} 
    and let  $P\in\mathbb{C}[x]$ be a degree-$k<n$ polynomial such that
    \begin{enumerate}
        \item P has parity $n$ mod 2
        \item $ \forall x\in[-1, 1]$, $\left|P(x)\right|\leq 1$
         \item $ \forall x\in(-\infty, -1] \cap [1, \infty)$, $\left|P(x)\right|\geq 1$
          \item if $n$ is even, then $\forall x\in\mathbb{R}$, $P(ix)P^*(ix)\geq 1$
    \end{enumerate}
    there exists $Q(x)\in \mathbb{C}[x], \Phi\in\mathbb{R}^{n+1}=\{\phi_0, \phi_1,...\phi_{n}\}\in\mathbb{R}^{n+1}$ such that
     \begin{align}\label{eq:Uphi_Gilyen_thrm3}
    U_{\phi}&=e^{i\phi_0\sigma_Z}\prod_{j=1}^n\left(W(x)e^{i\phi_j\sigma_z}\right)\\
    &=\begin{bmatrix}
        P(x) & iQ(x)\sqrt{1-x^2} \\
        iQ^*(x)\sqrt{1-x^2} & P^*(x)
    \end{bmatrix}.
\end{align}
\end{lemma}

An important restriction is when we solve for a real polynomial. Then,
\begin{lemma}[Symmetric QSP, para \cite{wang_energy_2022}]\label{lemma:qsp_symmetric}
    Given any $P\in\mathbb{C}[x], Q\in\mathbb{R}[x]$ such that 
    \begin{enumerate}
        \item $\text{deg}(P)=n$ and $\text{deg}(Q)=n-1$
        \item P has parity $n$ mod 2 and q has parity $n-1$ mod 2 
        \item $ \forall x\in[-1, 1]$, $\left|P(x)\right|^2+(1-x^2)\left|Q(x)\right|= 1$
         \item $ \forall x\in(-\infty, -1] \cap [1, \infty)$, $\left|P(x)\right|\geq 1$
          \item if $n$ is odd, then the leading coefficient of $Q$ is positive
    \end{enumerate}
    Then, there exists a unique set of symmetric phase factors 
    \begin{equation}
        \Phi_{d+1}\in \begin{cases}
            [-\pi/2, \pi/2)^{n/2}\times[-\pi, \pi)\times[-\pi/2, \pi/2)^{n/2},\\
            n \text{ is even}\\
             [-\pi/2, \pi/2)^{n+1} \\
             n \text{ is odd}
        \end{cases}
    \end{equation}
    such that
    \begin{equation}
        U(x, \Phi)=\begin{bmatrix}
            P(x) & iQ\sqrt{1-x^2}(x)\\
            i\sqrt{1-x^2}Q(x) & P^*(x)\\
        \end{bmatrix}
    \end{equation}
\end{lemma}
 Angle sets can be translated between conventions easily \cite{gilyen_quantum_2019, martyn_grand_nodate}. Furthermore, all angle-QSP circuits can be treated as special cases of Laurent-QSP or G-QSP circuits. To use QSVT, one must convert to the `reflection' convention \cite{gilyen_quantum_2019}, which modifies the signal operator. Gily{\'e}n et al. showed that the reflection QSP circuit can be extended into a circuit rotating the block-encoded singular values of a matrix \cite{gilyen_quantum_2019}. Herein, we will focus exclusively on symmetric-QSP in our discussion of angle-QSP since the strongest QSP-processing results work within it.

\subsubsection{Laurent-QSP}
The circuit we call Laurent-QSP is the original QSP circuit introduced by Low and Chuang in \cite{Low_2017}. We follow the formulation from Haah in  \cite{Haah2019product}, which emphasizes the reciprocality conditions on the real and purely imaginary components of the input polynomial  $\mathcal{P}(e^{i\theta})$. 
\begin{restatable}[Laurent QSP, para. from \cite{Haah2019product}]{lemma}{QSPHAAH}
\label{lemma:QSP_Haah}
    Consider $2\pi$-periodic function $\mathcal{P}(e^{i\theta})=\mathcal{A}(e^{i\theta})+i\mathcal{B}(e^{i\theta})$, where $\mathcal{A}(e^{i\theta})^2+\mathcal{B}(e^{i\theta})^2\leq1$ and $\mathcal{A, B}$ are real-on-circle, pure polynomials. Then, there exists a unique decomposition into $E_p$'s and a unitary 
    $F(e^{i\theta/2})=E_0E_{P_1}(e^{i\theta/2})...E_{P_{2n}}(e^{i\theta/2})$ such that 
    \begin{equation}
        \mathcal{P}(e^{i\theta})=\bra{+}E_0E_{P_1}(e^{i\theta/2})...E_{P_{2n}}(e^{i\theta/2})\ket{+}
    \end{equation}
\end{restatable}
With more specific assumptions on $\mathcal{A, B}$, we can sometimes derive angle-QSP circuits from Laurent-QSP \cite{Haah2019product, gilyen_quantum_2019}.

\subsubsection{Generalized-QSP}
The most general representation, G-QSP, comes from Motlagh and Wiebe in \cite{motlagh_generalized_2023}; it removes all restrictions on $P$ save the fairly benign requirement $|P(\cdot)|\leq 1$. If $|P(z)|\leq 1$, then we can find some set of complementary polynomials to embed $P$ in $SU(2)$ \cite{motlagh_generalized_2023}.
\begin{lemma}[G-QSP \cite{motlagh_generalized_2023}]\label{lemma:QSP_Motlagh}
    For all $n\in\mathbb{N}$, there exists $\Vec{\theta}, \Vec{\phi}\in\mathbb{R}^{d+1}$, $\lambda\in\mathbb{R}$ such that 
    \begin{equation}
        A_{\Vec{\theta}, \Vec{\phi}}=\left(\prod_{j=1}^dR(\theta_j, \phi_j, 0)A\right)R(\theta_0, \phi_0, \lambda)=\begin{bmatrix}
            {P}(U) & \cdot\\
            \cdot & \cdot\\
        \end{bmatrix}
    \end{equation}
    if and only if for all $z\in U(1)$, $\left|{P}(z)\right|^2\leq 1$. The $SU(2)$-rotation operator is defined as
    \begin{equation}
        R(\theta_j, \phi_j, \lambda)=\begin{bmatrix}
            e^{i(\lambda+\phi)}\cos\theta & e^{i\phi}\sin\theta \\
            e^{i\lambda}\sin\theta & -\cos\theta
        \end{bmatrix}\otimes I_U
    \end{equation}
\end{lemma}
Rotation gates $R$ are found with a recursive and efficient algorithm found in \cite{motlagh_generalized_2023}. Importantly, we can substitute oracle $CU+CU^{\dag}$ for $A$, then the circuit builds a degree $n$ Laurent polynomial\footnote{equivalently, we can derive a special case of G-QSP heuristically by replacing every $U^{\dag}$ gate in \cref{eq:QET_operator} with $U$, which builds a $2n$ degree complex polynomial instead of a degree $n$ Laurent polynomial.}. Presuming $CU+CU^{\dag}$ can be implemented with fewer than $2$ calls to $U, U^{\dag}$, then the circuit can, in some cases, achieve a factor of two improvement in query complexity, see \cite{berry2024doubling}. However, this circuit introduces a parity requirement on the complementary polynomial $Q$.
  
\section{QSP Circuits}\label{sec:circuits}
\subsection{block-encodings}
In general, one does not have access to a suitable oracle $U$ to implement QSP. Instead, the eigenvalues of $H$, or $\arccos(H)$ must be embedded within some unitary matrix. Then, one can modify the QSP circuit, rotating around the appropriate subspaces of this unitary at each step. \textit{qubitization} is the result from Low and Chuang in \cite{low_hamiltonian_2019} showing that such an oracle can be constructed for any Hermitian matrix. One must first construct a unitary $\Tilde{U}$, using some $s$-qubit state space such that $2^s\geq \text{dim}(H)$ and $a$ ancillary qubits, such that 
\begin{equation}
    \left(\bra{G}_a\otimes I_s\right)\tilde{U}\left(\ket{G}_a\otimes I_s\right)=H\label{eq:low_chuang_standardform}.
\end{equation}
$G$ represents a subspace of $SU(2)$ such that for any eigenvector $\ket{\lambda}$,
\begin{equation}
    \Tilde{U}\ket{G}\ket{\lambda}=\lambda\ket{G_{\lambda}}+\sqrt{1-|\lambda|^2}\ket{G_{\lambda}^{\perp}}.
\end{equation}
When $\ket{G}=\ket{0^{a}}$, \cref{eq:low_chuang_standardform} is our first example of a block encoding.
 \begin{definition}[Block encoding, paraphrased from \cite{gilyen_quantum_2019}]
    Suppose that $A\in\mathbb{C}^{n\times m}$, where $n, m\leq 2^s$ for $s\in\mathbb{Z}_+$. Then define embedding matrix $A_c\in\mathbb{C}^{2^s\times 2^s}$ such that the top-left block of $A_e$ is $A$, and all other elements are zero. This is a faithful embedding of matrices, producing $s$-qubit operator $A_c$. 
    Then, given $\alpha, \epsilon\in\mathbb{R}_+$ and $a\in\mathbb{N}$, we say that if $(s+a)$-qubit unitary $U$ obeys relation
    \begin{equation}
        \left|\left|A_e-\alpha\left(\bra{0}^{\otimes a}\otimes I_s\right)U\left(\ket{0}^{\otimes a}\otimes I_s\right)\right|\right|\leq \varepsilon
    \end{equation}
    then it is a block encoding of $A_c$. (and hence of $A$).
\end{definition}

Define subspaces $\mathcal{H}_{\lambda}=\text{span}\{\ket{G_{\lambda}}, \ket{G_{\lambda}^{\perp}}\}$, which are not invariant under the action of $\tilde{U}$. We require an iterate $W$ that block encodes $H$ as $\tilde{U}$ does above, and is also invariant about 2d-subspaces defined for each eigenvector, $\text{span}\{\ket{G_{\lambda}}, \ket{G_{\lambda}^{\perp}}\}$. Specifically,
\begin{align}
    W&=\bigoplus_{\lambda}\begin{bmatrix}
        \lambda & -\sqrt{1-|\lambda|^2}\\
        \sqrt{1-|\lambda|^2} & \lambda
    \end{bmatrix}_{\lambda}\\
    &=\bigoplus_{\lambda} e^{-i{Y}_{\lambda}\arccos(\lambda)}.
\end{align}
where $Y_{\lambda}$ is the operator such that $Y_{\lambda}\ket{G_{\lambda}}=i\ket{G_{\lambda}^{\perp}}$.From \cite{low_hamiltonian_2019}, we are guaranteed $W$ can be constructed for any $\Tilde{U}$ fulfilling \cref{eq:low_chuang_standardform}, with one query to $C\Tilde{U}$ and its inverse, one additional qubit, and $\mathcal{O}(\log(\text{dim}(\mathcal{H})))$ additional quantum gates.  Such a construction arises in discrete quantum walk oracles, where we can explicitly build oracles with eigenvalues $ie^{\mp i\arccos(\lambda)}=\pm e^{\pm i\arcsin(\lambda)}$ \cite{Childs_2009}. Very often, the construction of $W$ uses a linear combination of unitaries \cite{low_hamiltonian_2019} or a sparse access encoding.\\

 The term \textit{signal operator} is commonly used for whatever iterate is used in QSP/QET/QSVT. The \textit{signal processing operators} are the one qubit operations which rotate the signal ($R(\phi_i), V_{P_i}, R(\theta_i, \phi_i, \lambda)$). The use of block encodings/qubitized oracles requires that we promote the signal operator\footnote{QSVT literature usually promotes the signal processing operator instead \cite{gilyen_quantum_2019}.} into a controlled product of reflections about the subspace associated with $H$, that is $W=C(2\ket{G}\bra{G}-I)$ for the qubitized oracle. This promotion to qubitized operators defines the \textit{quantum eigenvalue transformation} (QET), where a polynomial is applied to the eigenvalues of some matrix $H$ encoded within $U$.\\

The QSVT case is analogous but now requires constructing unitary block encodings of matrices $A$, which, in general, can be non-square. Then, QSVT applies a polynomial $P$ to the singular values of matrix $A$ encoded within $U$. In recent years, methods have been developed for explicitly constructing block-encodings \cite{Zhang_2024, toyoizumi_hamiltonian_2023, sunderhauf_block-encoding_2023, camps_explicit_2023} or adapting QSP to Trotterized oracles \cite{silva_fourier-based_2022}. However, the block-encoding choice does not directly change the pre-processing, so it has also become common in QSP-processing literature to ignore the overhead costs of preparing block-encoded oracles. We omit further discussion of QET, QSVT, and their exact construction for space; readers are referred to \cite{martyn_grand_nodate, gilyen_quantum_2019} for a discussion in angle-QSP and \cite{sunderhauf2023generalizedquantumsingularvalue} for the extension for G-QSP. 

\subsection{LCU for QSP and QSVT circuits}
\label{sec:LCU_circuits}
 QSP circuits can easily be adapted into controlled QSP circuits by adding controls to the signal processing operators only - see \cref{sec:LCU_QSVT}.  Then, the signal operator (whether $CU, W$, or a block-encoding) can be `repurposed' for multiple QSP sequences at once, so long as the signal processing operations are controlled on distinct subspaces of the ancillary Hilbert space. Typically\footnote{ The most general approach is found in \cite{gilyen_quantum_2019} for QSVT block-encodings and \cite{laneve2023qsp} presents a similar technique for G-QSP}  one prepares an $m$-qubit ancillary space in $H^{\otimes m}\ket{0^{\otimes m}}$, and then encodes up to $2^m$ length-$n$ QSP-operations at the cost of subnormalization $\frac{1}{2^m}$ and up to $m+1$ ancillary qubits. We reproduce a version here, \cref{thrm:qsplcu}, which is stated for G-QSP so that it will automatically apply to every QSP convention. 
\begin{restatable}[LCU for QSP]{theorem}{LCUQSP}\label{thrm:qsplcu}
    Consider a set of $L_e$ G-QSP valid polynomials $\{h_i\}$ with even degree $n$, and a set of $L_o$ G-QSP valid polynomials $\{g_i\}$ with odd degree $n+1$. Define $l=\lceil\log_2(L_e+L_o)\rceil$, $l_o=\lceil\log_2(L_o)\rceil$  and assume a signal operator $U$ and set of signal processing operators $\{R\}$. We can implement the following polynomial transformation 
    $$F(e^{i\theta})=\frac{1}{2^l}\left(\sum_{i=1}^{L_e} h_i(e^{i\theta})+\sum_{i=1}^{L_o} g_i(e^{i\theta})\right)$$
    on iterate $U$, on the general form of a QSP circuit given in  \cref{fig:QET_basic_circuit} and the following cost
    \begin{align}
     \cost{F(U)}_{QSP} &=\left(L_e(n+1)+L_o(n+1)\right)\cost{C_{l}R}\nonumber\\
     &+2l\cost{H}+n\cost{U}+L_o\cost{C_{l_o}R}\nonumber\\
     &+l_o\cost{C_{l_o}U} 
    \end{align}
    where $\cost{\cdot}$  denotes the cost of a gate.
    Besides the state space of $U$, the protocol uses $l+1$ ancillary qubits when $h_j, g_k$ are complex polynomials $\forall j, k\in 1,2...l$, and $l+2$ ancillary qubits when $h_j, g_k$ are real $\forall j, k\in 1,2...l$.
\end{restatable}
 We give a constructive proof of the result in \cref{sec:LCU_QSVT}. 
The notation $\cost{G}$ can be used for any standard measure of the cost of gate $G$ - for example, the number of $T$ gates used in preparing $G$ or the gate complexity of $G$.

\subsection{Comparative Query Complexity of QSP}\label{sec:comp_query_complex}
 Here we reproduce the gate complexity of preparing complex, degree $n_{max}$ polynomial $\mathcal{A}$ such that $|\mathcal{A}|\leq 1/2$ for all $z\in U(1)$ in each convention. Without loss of generality, we assume $n_{max}$ is odd. In this section, we assume all QSP-oracle costs are analogous.
 We assume the initial state has the following decomposition in the eigenbasis of $U$, $\ket{\psi}=\sum_i\psi_i\ket{\lambda_i}$. Then, the post-selected state has the form $\Tilde{\Pi}U_{QSP}\ket{\psi}=\frac{1}{\alpha}\sum_i \psi_if(\lambda_i)\lambda_i$. Cumulatively, the subnormalization $\alpha$ arises from the block-encoding and any subnormalization in the QSP-processing. Altogether, $p_{succ}=\alpha^{-2}{\left|\left|\sum_i \psi_if(\lambda_i)\lambda_i\right|\right|^2}$.
Oblivious quantum amplitude amplification from \cite{gilyen_quantum_2019} can be used to boost the success probability of any of the following, at the cost of a query overhead $\mathcal{O}(\frac{1}{\sqrt{p_{suc}}})$.\\

\subsubsection{Generalized-QSP Gate Complexity}
Generalized QSP in \cite{motlagh_generalized_2023} has the shortest gate complexity for a generic $\mathcal{A}(z)$, featuring the fewest controlled operations and using only one ancillary qubit. We simply have
\begin{equation}\label{eq:gqsp_query_complex_U}
    \mathcal{C}[U_{GQSP, \mathcal{A}}]= (n_{\max}+1)\cost{CU}+\left(n_{max}+1\right)\cost{R}.
\end{equation}
The success probability is $p_{succ}=\frac{\left|\left|\sum_i \psi_if(\lambda_i)\lambda_i\right|\right|}{\alpha}$. 

\subsubsection{Angle-QSP Gate Complexity}
Using angle-QSP, we use the conversion $P(z)\rightarrow \sum f_{i, j}(x)$ defined in \cref{eq:poly4termdecomp} and then go through $4$ QSP-processing steps within precision $\epsilon_i$, one for each $f_{i, j}$. We require four ancilla qubits\textendash the QSP ancilla, an ancilla qubit to extract the real or purely imaginary part $f_{i, j}$ embedded in a complex QSP polynomial, and then two ancilla qubits for the LCU step. Controlled phase gates $\text{Ph}(\frac{\pi}{2})$ add the appropriate factors of $i$. Assume every $f_i$ has a degree bound by $n$ and is approximated within precision $\epsilon$. The QSP (or QSVT) approximation to $\mathcal{F}(z)$ has the following gate complexity
\begin{align}\label{eq:angle_query_complex_U}
    \cost{U_{angle, P}}&=\left(\sum_{i=0}^3\cost{U_{angle, f_i}}+6\cost{H}+2\cost{C\text{Ph}(\frac{\pi}{2})}\right)\\
&=\mathcal{O}\left(n_{\max}\cost{W}+2\cost{C_0W}\right).
\end{align}
where each distinct QSP circuit is
\begin{equation}
    \cost{U_{angle, f_i}}=\begin{cases}
         2(n_{i}+1)\cost{e^{i\sigma_j}}+2\cost{H}
        +(n_{i}-1)\cost{W} +\cost{C_0W} & n_i \text{ odd}\\ 
        n_{i}\cost{W} + 2(n_{i}+1)\cost{e^{i\sigma_j}}
        +\cost{C_0W} +2\cost{H}& n_i \text{ even}\\ 
    \end{cases}
\end{equation}
\
The success probability is  $p_{succ}=\frac{\left|\left|\sum_i \psi_if(\lambda_i)\lambda_i\right|\right|^2}{16\alpha^2}$.

\subsubsection{Laurent-QSP Gate Complexity}
If we instead use the Laurent-QSP framework, then we can use two circuits\textemdash respectively for $\mathcal{F}(z)=f_{R,E}(x\rightarrow z)+if_{I, O}(x\rightarrow z)$ and $\mathcal{G}(z)=i\left(f_{I,E}(x\rightarrow z)-f_{R, O}(x\rightarrow z)\right)$ implemented with the help of a controlled phase gate $C\text{Ph}(\frac{\pi}{2})$. Each $\mathcal{F}(z)$ is now in Laurent polynomial form, so along with being real-on-circle, it is reciprocal or anti-reciprocal and hence fits the conditions in \cref{lemma:QSP_Haah}, at the expense of a factor of $2$ increase in the circuit length. The Laurent-QSP circuit has the following gate complexity
\begin{align}
    \mathcal{C}[U_{QSP, f_i}]&=2n_{nmax}\cost{CU}+2\cost{H}+\cost{C\text{Ph}(\frac{\pi}{2})}+2\left(2n_{max}+1 \right)\cost{V}.
\end{align}
This approach requires two ancillary qubits and has success probability $p_{succ}=\frac{\left|\left|\sum_i \psi_if(\lambda_i)\lambda_i\right|\right|^2}{4\alpha^2}$. The increased depth may be, in some cases, preferable to the reduced circuit length and additional ancilla qubits in the symmetric-QSP case.

\section{Fej{\'e}r-Wilson QSP-Processing}\label{sec:fejer}
We now present one QSP-processing method in detail, focusing on the completion step. Details of the decomposition step can be found in appendix \ref{sec:Haah_proof}.

\subsection{The Fej{\'e}r Problem}
We begin by observing that the Laurent-QSP convention of \cref{lemma:QSP_Haah},  \cref{prob:qsp_poly_constraint} can be formulated as a known problem in complex analysis.
\begin{definition}[Fej{\'e}r Factorization, from \cite{goodman_spectral_1997}]\label{def:fejer_fact}
Consider Laurent polynomial $\mathcal{F}(z)=\sum_{k=-n}^nF_kz^k$ such that the coefficients satisfy $F_i=F^*_i=F_{-i}$, and $\mathcal{F}(e^{i\theta})\geq 0$ for $\theta\in\mathbb{R}$. Find a sequence of real numbers $\{\gamma_0, \gamma_1, ....\gamma_n\}$ defining polynomial $\gamma(z)=\sum_{i=0}^{n}\gamma_iz^i$, $\gamma(0)>0$, such that
\begin{equation}
    \mathcal{F}(z)=\gamma(z)\gamma(1/z)
\end{equation}
and $\gamma$ has all roots outside $U(1)$.
\end{definition}
Constraining our solution to have all roots greater than unity and requiring $\gamma(0)>1$ guarantees a unique solution for $\gamma$. There are several methods for solving the Fe{\'j}er problem, of which two methods with bounded error outperform root-finding on the benchmarks in \cite{goodman_spectral_1997}. We selected the (usually) quadratically convergent Wilson method from \cite{wilson_factorization_1969} for QSP-processing. 

\subsection{Wilson Method}\label{sec:wilson}
We assume access to the coefficients of $\mathcal{F}$, $\{F_i\}$, and an acceptable precision, $\epsilon_{fejer}$. Coefficients of $\mathcal{F}$ obey
\begin{equation}\label{eq:wilson_coeff_sum}
   F_i=\sum_{j=0}^{n-i}\gamma_j\gamma_{j+i}.
\end{equation}
Then, we can construct a system of equations for the coefficients that will uniquely specify $\gamma$, and since $F_i=F_{-i}$ for all $i$, we have only $n+1$ distinct coefficients. By solving $\sum_{j=0}^{n-i}\gamma^l_j\gamma^l_{j+i}-F_l$ with Newton-Raphson iteration and a suitable initial guess $\gamma^{0}$, one identifies a unique choice of $\gamma$ within $\epsilon_{fejer}$.  First construct vectors 
\begin{equation}
\vec{F}=\begin{bmatrix}
    F_0\\
    F_1\\
    F_2\\
    \vdots\\
    F_n
\end{bmatrix}    \quad
\vec{\gamma}=\begin{bmatrix}
    \gamma_0\\
    \gamma_1\\
    \gamma_2\\
    \vdots\\
    \gamma_n
\end{bmatrix}   
\end{equation}
$\partial_kF_i=\gamma_{k+i}+\gamma_{k-i}$, so define triangular matrices 
\begin{equation}
    T_1=[\gamma_{i+j}]_{ij} \quad T_2=[\gamma_{j-i}]_{ij} \quad  i, j=0,1...n
\end{equation}
 where for $ i+j>n, \:[\gamma_{i+j}]_{ij}=0.$ It is convenient to define the vector form of $ \sum_{j=0}^{n-i}\gamma^l_j\gamma^l_{j+i}$,
 \begin{equation}
    \Vec{c}^l=T_1^l\gamma^l=T^l_2\gamma^l.
 \end{equation}
Then Newton-Raphson iteration step is equivalent to solving the following system of linear equations
\begin{equation}
    (T_1+T_2)\gamma=\Vec{c}+\vec{F},
\end{equation}
which can be solved with standard linear solvers. In this work's benchmarks, NumPy's linalg.solve function is sufficient. From \cite{goodman_spectral_1997, wilson_factorization_1969}, if we require the initial guess $\gamma^0$ to have all zeros outside of the unit circle, we are guaranteed that all subsequent steps $\gamma^k$ have all zeros outside the unit circle\footnote{See  \cite{goodman_spectral_1997} for an alternate choice which may sometimes have a stability advantage.}. One suitable initial guess is to use constant function $\gamma^{0}(z)=\gamma_0^0$ for some $\gamma^0_0>0$ \cite{wilson_factorization_1969}. 

\begin{algorithm}
    \begin{algorithmic}
        \Procedure{Wilson Iteration}{$F, n, \epsilon_{fejer}$}
\Comment Coefficient list $F$, degree $n$, initial guess $\gamma_0$, and solution precision $\epsilon_{qsp}$
\State Build vector $c$ and matrices $T_1, T_2,$ from  $\gamma^{(0)}$ 
\State Solve $(T_1+T_2)\gamma'=c+F$, and then define $\gamma^{1}=\frac{1}{2}\gamma'-\delta$
\While {$\linfnorm{\gamma^{(j)}-\gamma^{(j-1)}}\geq \epsilon_{fejer}$ and $j\in\{0, 1, 2, ...n^2\}$}
\State Build vector $c$ and matrices $T_1^{(j)}, T_2^{(j)}$ from  $\gamma^{(j)}$  
\State Solve $(T_1^{(j)}+T_2^{(j)})\gamma'=c^{(j)}+F$, and then the solution is $\gamma^{(j+1)}$
\EndWhile
\State \textbf{Return} $\tilde{\gamma}$
\EndProcedure
    \end{algorithmic}
    \caption{Wilson Method}
\label[algorithm]{alg:wilsonalg}
\end{algorithm}

\begin{figure}[ht!]
    \centering
    \input{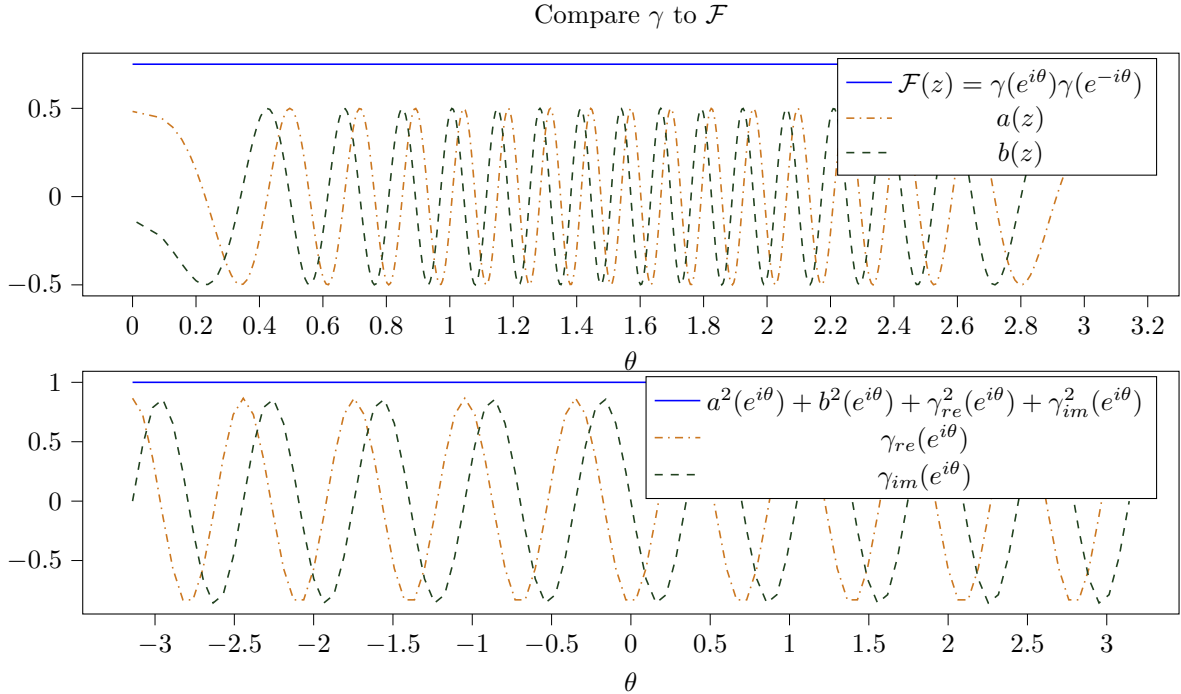}
    \caption{The Wilson method for sample $\mathcal{F}(z)$ and $\gamma(z)$ solution. Above: real-on-circle polynomials $a, b$ are used to define Laurent polynomial $\mathcal{F}=1-a^2(z)-b^2(z)$ which fits the criteria in \cref{def:fejer_fact} to be factorized. The solution $F(z)=\gamma(z)\gamma(z^{-1})$ is in this case constant. Below: the real and imaginary parts of the solution $\gamma$, along with confirmation that constraint $a^2(z)+b^2(z)+\gamma_{re}^2(z)+\gamma^2_{im}(z)=1$. $a, b$ are defined from a Jacobi-Anger expansion with $n=90$ and sub-normalization $\frac{1}{\sqrt{2}}$, see \cref{sec:hs_poly}.}
    \label{fig:gamma-demo}
\end{figure}

\begin{lemma}[Wilson Method]\label{lemma:wilson}
    Given a reciprocal, real-on-circle, and real Laurent polynomial  $\mathcal{F}$, its coefficient set $\{F_i\}$, and precision $\epsilon\in (0, 1)$, \cref{alg:wilsonalg} prepares $\tilde{\gamma}$ such that 
    $$\linfnorm{\tilde{\gamma}-\gamma}\leq \epsilon$$
    where $\gamma$ is the solution to the Fej{\'e}r problem, so that $\mathcal{F}(z)=\gamma(z)\gamma(z^{-1})$.
\end{lemma}
The proof is found in \cite{wilson_factorization_1969}. In \cref{fig:gamma-demo}, we apply the Wilson method to a simple test polynomial. 

\subsubsection{Computational Complexity}
We have so far disregarded the computational complexity of the Wilson method. The dominant subroutine is solving a system of linear equations, which may scale with $\mathcal{O}(n^3)$  floating point operations in the worst case. The complexity also depends on the number of iterations and, hence, on the convergence rate of the Newton-Raphson subroutine discussed in \cref{sec:wilson-convergence}. In \cite{wilson_factorization_1969, goodman_spectral_1997}, the Wilson method is expected to converge quadratically, and so the costs of the Wilson method scale with $\mathcal{O}(n^3\log(1/\epsilon))$. However, the convergence rate actually depends on the matrix norm of the inverse of the Jacobean of $\mathcal{F}$. In \cref{lemma:wilson-convergence}, we show that, with additional criteria on the input polynomial, there exists an initial guess such that better than linear convergence is always possible. Unfortunately, this guess is not trivial to identify.

\begin{restatable}[better than linear convergence of the Wilson method]{lemma}{WILSONCONV}\label{lemma:wilson-convergence}
    Given $\delta\in(0, 1]$ and input Laurent polynomial $\mathcal{F}(z)$ with truncated coefficient list $(F_0, F_1,...F_{n})$, whenever the ideal solution $\gamma^{(*)}$ obeys
    \begin{align}\label{eq:wilson-condit-eq}
  \frac{|\gamma^{(*)}_J|}{\left|\left|(T)\right|\right|_r}&\leq \frac{B}{(n+1)^2}, \quad B=\frac{1-\delta}{3-2\delta}
\end{align}
then there exists an initial guess $\gamma_0$ such that \cref{alg:wilsonalg} will converge with rate at least $1-\delta$ to solution with error at most $(n+1)\epsilon$ for some $\epsilon\in \left[0, \frac{1}{(n+1)}\right)$.
\end{restatable}
The proof is given in \cref{sec:wilson-convergence} but is unfortunately not constructive. That is, without  a bound on the coefficient of $\gamma^{(*)}$, we cannot use \cref{lemma:wilson-convergence} to select an adequate initial guess or check if the function has the correct properties to converge. 

\subsection{Fej{\'e}r-Wilson method}
\begin{algorithm}
\begin{algorithmic}
\Procedure{QSP-Processing}{$\mathcal{\Tilde{A}, \Tilde{B}}, n, c, \epsilon_{QSP}$}
\Comment Coefficient lists $\mathcal{\Tilde{A}, \tilde{B}}$ , degree $n$, solution precision $\epsilon_{qsp},$ constant $c$
\If {$\mathcal{\Tilde{A}, \tilde{B}}$ take arguments on $x\in \left[-1, 1\right]$,} 
  \State convert to length $2n+1$ Laurent polynomial lists $\mathcal{\Tilde{A}, \tilde{B}}$.
\EndIf
\If {$\mathcal{\Tilde{A}}=0$ or $\mathcal{\Tilde{B}}=0$ } 
   \State redefine $\mathcal{\Tilde{A}}$ or $\mathcal{\Tilde{B}}$ as a degree $n-1$ sum of Chebyshev polynomials with random coefficients.
\EndIf
\State Build coefficient list for $1-\mathcal{\Tilde{A}}^2(z)-\mathcal{\Tilde{B}}^2(z)$ 
\State Solve the Fej{\'e}r problem for $1-\mathcal{\Tilde{A}}^2(z)-\mathcal{\Tilde{B}}^2(z)$ using the Wilson method, obtaining coefficient list $\gamma$
\State Build coefficient lists for $\mathcal{\Tilde{C}, \Tilde{D}}$ from  $\gamma$
\State Build matrix-valued coefficient list for $$F(z)=\mathcal{\Tilde{A}}(z)I+i\mathcal{\Tilde{B}}(z)\sigma_X+i\mathcal{\Tilde{C}}(z)\sigma_Y+i\mathcal{\Tilde{D}}(z)\sigma_Z$$
\State Convert the coefficient list of $F(z)$ to the coefficient list of $F^{(2n)}(z^2)=F(z)$ on $z\in U(1)$ by padding with zeros.
\While{$F^{j}(z)$ has degree $j>1$}
\State Compute near-projector $P$ from $C_{j}^{j}$
\State Compute eigenvectors $\{\ket{e_0}\ket{e_1}\}$ of $P$, where $e_1\propto \epsilon_{qsp}$, and define projector $\Tilde{P}^{(j)}=\ket{e_0}\bra{e_0}$
\State Define unitary $E_{\tilde{P}^{j}}$
\State Compute the coefficients of $F^{(j-1)}$ from $E_{\tilde{P}^{j}}, F^{(j)}$
\EndWhile
\State Set $F^{(1)}=E_0$
\State \textbf{Return} $E_0, \{\Tilde{P}^{(j)}\}$
\EndProcedure
\end{algorithmic}
\caption{QSP-processing with the Wilson Method}
\label[algorithm]{alg:fejer_qsp_processing}
\end{algorithm}
Proving that a QSP-processing method based on the Wilson method will succeed with bound error requires a few conditions in addition to those in \cref{lemma:QSP_Haah}. The algorithm whose error we will bound is \cref{alg:fejer_qsp_processing} from \cite{skelton2024harmlessmethodsqspprocessinglaurent} and modeled after Haah's method in \cite{Haah2019product}, which is discussed in more detail in \cref{sec:haah-method-summary}.\\

The  Fej{\'e}r-Riesz theorem justifies the existence of a polynomial $\gamma(z)$, which solves the completion step of QSP-processing. From \cref{lemma:wilson-convergence}, we have identified conditions under which \cref{alg:wilsonalg} can solve it. Unfortunately, the condition requires prior knowledge of the coefficients of $\gamma$ and so cannot be used to check if a given polynomial really fits the condition. However, QSP processing methods are often used successfully outside their convergence conditions, and we similarly defer to \cref{sec:results} to demonstrate that the completion step is normally successful for QSP-relevant benchmark polynomials.\\

After the completion step, one must manage how error propagates through the decomposition step of QSP-processing. Error in the polynomials $\Tilde{\mathcal{C}(z)}, \Tilde{\mathcal{D}(z)}$ identified in the decomposition step carries through, leading to error in the projectors $\{\Tilde{P}\}$ defining the QSP circuit. \Cref{alg:fejer_qsp_processing} modifies the decomposition step proposed in \cite{Haah2019product} to ensure that the resulting QSP parameters $\{\Tilde{P}\}$ form unitary operators.  We now present the novel theorem of this work, which bounds the error of QSP-processing with the Wilson method. \\
\begin{restatable}[Laurent QSP solved with the Wilson Method]{theorem}{LQSPTHRM}\label{thrm:laurent_qsp_with_processing}
    Consider $2\pi$-periodic function $\mathcal{P}(e^{i\theta})=\mathcal{\Tilde{A}}(e^{i\theta})+i\mathcal{\Tilde{B}}(e^{i\theta})$, which is $\frac{\epsilon}{2}$-close to target function $f$  for $\epsilon\in [0, 1)$, such that the following conditions are met
    \begin{enumerate}
    \item $\linfnorm{\mathcal{P}}\leq1$, $\mathcal{\tilde{A}, \tilde{B}}$ are real-on-circle, pure polynomials 
        \item coefficients of $\mathcal{\Tilde{A}}, \mathcal{\Tilde{B}}$ can be encoded to accuracy $\frac{\epsilon}{2(2n+1)}$
        \item the leading coefficients satisfy  $|\mathcal{\Tilde{A}}_{n}|, |\mathcal{\Tilde{B}}_{n}|, |\mathcal{\Tilde{A}}_{-n}|, |\mathcal{\Tilde{B}}_{-n}|\geq \frac{c}{2}$ for some $c>0$ 
        \item conditions of \cref{lemma:wilson-convergence} are met for $1-|\mathcal{P}(e^{i\theta})|^2$.
    \end{enumerate}
    We select precision $\epsilon_{fejer}\leq\frac{\epsilon}{2c_n}$ for the Fej{\'e}r method, with constant $c_n=\frac{4(2n+1)}{c}2^{2n}$. Then, there exists a decomposition into $E_p$'s and a unitary 
    $F(e^{i\theta/2})=E_0E_{P_1}(e^{i\theta/2})...E_{P_{2n}}(e^{i\theta/2})$ such that 
    \begin{equation}
        \left|\mathcal{P}(e^{i\theta})-\bra{+}E_0E_{P_1}(e^{i\theta/2})...E_{P_{2n}}(e^{i\theta/2})\ket{+}\right|\leq \epsilon.
    \end{equation}
\end{restatable}
In \cite{Haah2019product}, arbitrary precision arithmetic is used to bound the error and numeric instabilities in the completion and decomposition steps. In our presentation, encoding coefficients to the desired precision may require one to work outside of floating point precision arithmetic. Note that if one wants to instead prepare $\mathcal{P}_{real}=\mathcal{\Tilde{A}}$ or  $i\mathcal{P}_{imag}=i\mathcal{\Tilde{B}}$, one can randomly select a complimentary $\mathcal{\Tilde{B}}, \mathcal{\Tilde{A}}$ and then procede with \cref{thrm:laurent_qsp_with_processing}. One then extracts only the real or imaginary component of the resulting QSP circuit using the standard LCU techniques discussed in \cref{sec:circuits}.

\section{QSP-processing Methods}\label{sec:qsp_processing}
We review current and historic strategies for solving QSP-processing. We do not give explicit algorithms, nor do we go into extensive mathematical detain on any method. The aim is instead to sketch out the strategy, dominant subroutines, and advantages and drawbacks of each method. Note that the numeric results discussed below are not exactly comparable due to differences between each implementation's polynomial approximation methods. 

\subsection{Direct QSP-Processing}
Direct methods improve the constructive proofs found in Gily{\'e}n et al. \cite{gilyen_quantum_2019} for angle-QSP. Given polynomial $P: \mathbb{R}\rightarrow \mathbb{C}$ fitting the requirements of \cref{lemma:QSP_gilyen}, \cref{prob:qsp_poly_constraint} is $A(x)=1-|P(x)|^2$. The conditions on $P$ in \cref{lemma:QSP_gilyen} ensure that $A(x^2)$ has real roots with even multiplicity and complex roots in reciprocal pairs. Thus, knowing the roots of $A$, we can easily define a subset  $S$ of the root multiset such that
\begin{align}
    A(x^2)&=K^2(1-x^2)\prod_{s\in S}(x^2-s)(x^2-s^*)\nonumber\\
    &=(1-x^2)W(x^2)W^*(x^2)
\end{align}
for 
\begin{equation}
    W(x^2) \coloneqq K\prod_{s\in S}(x^2-s).
\end{equation}
We then set $Q(x)=\sqrt{1-x^2}W(x^2)$, resolving the \textit{completion step}. For all direct methods, the completion step is expected to be the dominant difficulty in finding QSP parameters.

Now, we have suitable degree $\leq n$ polynomials $P=P_n, Q=Q_n$ For the subsequent \textit{decomposition step},  \cite{gilyen_quantum_2019} gives a recursive equation for $\phi_k$ and degree $k-1$ polynomials $P_{k-1}(x), Q_{k-1}(x)$ with degrees respectively $\text{deg}(P_{k-1})\leq \text{deg}(P_k)-1$, $\text{deg}(Q_{k-1})\leq \text{deg}(Q_{k})-1$,  and $k=0,1,...n$. Unambiguous improvements to both steps are discussed below.

\subsubsection{Numerically stable root-finding}\label{sec:haah-method-summary}
In  \cite{Haah2019product}, Haah introduced a numerically alternative alternative to  Gily{\'e}n et al.\cite{gilyen_quantum_2019} by working in arbitrary float precision and altering the decomposition step. This work also explicitly shows how errors in the input polynomial and the completion step propagate through the decomposition step.

Recall that in the convention of \cref{lemma:QSP_Haah}, one works with a subset of Laurent polynomials $\mathcal{P}(z)$. All roots of \cref{prob:qsp_poly_constraint} in \cref{lemma:QSP_Haah} are complex reciprocal pairs, and none are on the unit circle \cite{Haah2019product}. In \cite{Haah2019product}, solution $\gamma$ is constructed from the root-set of the problem polynomial, and root-finding routines are the primary computational bottleneck. One obtains $\gamma(z)$ such that $Q(z)=\gamma(z)\gamma(z^{-1})$, then, the real and imaginary parts of $\gamma$ are used to define $\Tilde{\mathcal{C}}(z)=\mathcal{C}(z)+\mathcal{O}(\epsilon_{coeff})$, $\Tilde{\mathcal{D}}(z)=\mathcal{D}(z)+\mathcal{O}(\epsilon_{coeff})$.\\

Once the completion step is solved, one builds $\mathcal{F}(z)=\sum_iC_iz^i\in SU(2)$ from $\mathcal{P}(z), \mathcal{Q}(z)$. The coefficients of $\mathcal{F}(z)$ are extracted from a list of function values using a discrete fast Fourier transform (FFT). The decomposition step is completed by an iterative algorithm, \textit{carving} the matrix-valued coefficients of $F(z)$ into projector set $\{P_j\}$ and unitary $E_0$, which defines the QSP circuit. When we build $F(z)$, 
\begin{equation}
    \left[\Tilde{F}(z)\right]_{ij}=\left[{F}(z)\right]_{ij}+\mathcal{O}\left(\epsilon_{coeff}\right)
\end{equation}
The carving process results in the circuit
\begin{equation}
    F(z)=E_{0}\prod_{j=1}^{2n}E_{j}(\sqrt{z})+\varepsilon_{qsp}.
\end{equation}
The precision of carving is bounded when the precision of the completion step is bounded. However, due to error propagation in the decomposition step, this requires $\mathcal{O}(\text{poly}(n\log(1/\epsilon))$ digits of precision.

\subsubsection{Halving and Capitalization}\label{sec:DM_halving}
Working instead with the QSP convention of \cref{lemma:QSP_gilyen}, Chao et al. \cite{chao2020finding} replaced the carving procedure in \cite{Haah2019product}. They also introduced \textit{capitalization}, adding $\varepsilon$ order leading terms to the polynomial. This is an ad hoc solution to deal with very small coefficients in the leading terms \cite{chao2020finding}.\\

\textit{Halving} can mitigate the effect of numeric instability in the completion step. The {halving} algorithm is a recursive algorithm to split up $F(z)$ into a set of $n$ rotations about  $\sigma_z$. Given $F(z)=\sum_{j=-n}^nC_jz^j$ with matrix-valued coefficients $C_j\in M^{2x2}$, we can encode the coefficients of $F(z)$ into a block banded Toplitz matrix of dimension $4n\times 2(n+1)$, $l=1,...n-1$. Then one constructs a system of linear equations whose solution defines coefficient matrices $V_l(z)=\sum_{j=-l}^lC_jz^j$. One then defines $V_l^*(z)F(Z)=V_{n-l}$ at each step. A least squares solution for $V$ is used at each step in order to mitigate the effects of error carried through in the completion step. The error in Chao et al. relies on the success of the least squares solution as well as the completion step \cite{chao2020finding}. However, the resulting unitaries $\{V\}$ are guaranteed to be Pauli-Z rotations, and so one can compute angle set $\Phi=\{\phi\}$ from $\{V\}$\\

For Hamiltonian simulation with $\epsilon\in [10^{-2}, 10^{-6}], \tau\leq 1200$, halving allowed Chao et al. to identify solutions within a larger region of the parameter space than carving \cite{chao2020finding}. An existing open-source implementation using this method is used for numeric results in \cite{martyn_grand_nodate}. However, these methods still struggle with instances that the Prony and optimization methods discussed below can solve. 

\subsubsection{Prony Method}
A technique using Prony's method as a subroutine ostensibly works in the convention of \cref{lemma:QSP_Haah}. However, a key point of the technique is to engineer $\mathcal{B}(z)$ for a given $\mathcal{A}(z)$ so that only the real part of $\mathcal{P}(z)$ is the desired QSP polynomial. Thus, the original Prony method is only successful for symmetric-QSP instances. The method also requires $\mathcal{A}(z)$ bound well below one\textemdash in \cite{Ying2022stablefactorization}, $|\mathcal{A}(z)|, |\mathcal{B}(z)|<1/3$. 
$\mathcal{B}(z)$ is chosen as 
\begin{equation}
    \mathcal{B}(\theta)=b_d\sin(d\theta)+b_{d-2}\sin((d-2)\theta)+...
\end{equation}
with the condition that $b_d$ is the dominant coefficient and the rest are chosen randomly\cite{Ying2022stablefactorization}.\\

The Prony method is designed to bypass the difficult root-finding step in \cref{prob:qsp_poly_constraint}. Essentially, one uses a FFT to approximate contour integrals related to $g(z)=\left(1-\mathcal{A}^2(z)-\mathcal{B}^2(z)\right)^{-1}$. These contour integrals define (potentially noisy) data points in a series whose solution can be computed using Prony's method. This series solution produces the coefficients of $z^n\gamma(z)$.\\

The core routines used are the FFT and a conjugate gradient method; these can be provably bound for sufficiently restrictive conditions, although numeric work in \cite{Ying2022stablefactorization} shows the method can still converge to solutions with reasonable precision when these conditions do not hold. Phases are identified using a decomposition step most similar to the original result\cite{gilyen_quantum_2019}, and the precision is determined numerically as the $l^{\infty}$-norm between $\mathcal{P}$ and the resulting QSP circuit. Solutions with precision ranging from $\mathcal{O}(10^{-13}, 10^{-11})$ are achieved for similar test problems as considered here.\\

Finally, we note that Yamamoto and Yoshioka have extended the Prony method technique to complex polynomials  \cite{yamamoto2024robust}. Their published benchmark is Hamiltonian simulation for $t=10$ and $\epsilon_{qsp}=10^{-13}$. Additionally, they compare their Prony method implementation to the optimization method discussed in \cref{sec:brute_opt}. They found that optimization struggles for Hamiltonian simulation, requiring $\mathcal{O}(10^2)$ seconds for instances which the Prony method can handle in $\mathcal{O}(1)$. However, all instances for this method remain quite small in contrast to the benchmarks done for symmetric-QSP; the normalization they require is not stated, and the propagation of error through their technique is not explicitly given.

\subsubsection{BerS{\"u}n Method}
Berntson and S{\"u}nderhauf introduced the most general direct method for QSP-processing, suitable for all complex polynomials for G-QSP \cite{berntson2024complementarypolynomialsquantumsignal}. They also provide a full error analysis of the completion step of their method. The core technique is to work with the contour integral representation of the complimentary polynomial Q. On the unit circle, this contour integral is
\begin{equation}
    Q(e^{i\theta})=Q_0(e^{i\theta})e^{\prod\left[\log\left(\frac{1-|P(e^{i\theta})|^2}{|Q_0(e^{i\theta})|^2}\right)\right]}.
\end{equation}
where $\prod$ is the Fourier Multiplier and $Q_0(e^{i\theta})=\prod_{\{t_j\}}(z-t_j)^{\alpha_j}$ for $t_j$ is a root on the unit circle with multiplicity $2\alpha_j$.  This restriction to a canonical form of $Q$ ensures a unique solution up to a global phase.
$Q_0(e^{i\theta})=1$ for any $P$ such that $\linfnorm{P}<1-\delta$, $0\leq \delta$. This translates into a subnormalization condition on $P$, and we will assume from this point that $Q_0=1$. Note that for any polynomial fitting the conditions of \cref{lemma:QSP_Haah}, this condition holds automatically. Crucially, some of their benchmarks have $l^{\infty}$ norm $\epsilon$ close to one, whereas the above direct QSP-processing methods rely on $l^{\infty}$ norm at most $\frac{1}{\sqrt{2}}$ to succeed.\\

Written with respect to the Fourier expansion of $1-|P(e^{i\theta})|^2$ in the Laurent basis, $S(z)=\sum_{-n}^n a_jz^j$, we obtain an expression for $ Q(e^{i\theta})$ which depends only on $aj$, the coefficients of $S$. From this point, the algorithm is relatively simple. From the coefficients of $P\in \mathbb{C}$, one can compute $P(W)$ at $N+1$ roots of unity $W$ using an inverse FFT. Another FFT is used to approximate the Fourier coefficients of $\log\left(1-|P(e^{i\theta})|^2\right)$, and then approximate $Q(w)$. Polynomial interpolation ensures that $Q(w)$ produces a trigonometric polynomial, and the coefficients of $Q$ are extracted by a final FFT. The accuracy of the approximation will depend on some $N\in \mathbb{Z}_{\geq n}$, and \cite{berntson2024complementarypolynomialsquantumsignal} shows that for a given $\epsilon, \delta, n$, $N=\mathcal{O}\left(\frac{d}{\delta}\log\left(\frac{d}{\delta\epsilon}\right)\right)$. The method's complexity scales with $N\log(N)$, and succeeds for some high precision, $\mathcal{O}(10^{-16}, 10^{-10})$, and high degree, $10^{7}$, instances.

\subsection{Optimization Methods}
 \subsubsection{Symmetric-QSP}\label{sec:symmetric-qsp-processing}
 Because it is so difficult to identify the root structure of a high-degree polynomial, Dong et al. turned the identification of $Q(\cdot)$ into an optimization problem \cite{dong_efficient_2021}. When a solution is needed for a definite parity, real polynomial, one can assume $Q[x]\in\mathbb{R}[x]$ and introduce a symmetry requirement on the angle set\footnote{One might also try to solve this optimization problem for complex $P$. This is done successfully for the problem of Hamiltonian simulation in the supplementary information in \cite{Dong_2022}. However, the cost landscape of the complex case is not as well studied, so we do not discuss it here.}. The resulting simplification of QSP is symmetric-QSP.
 
\begin{problem}[Symmetric-QSP parameter identification, paraphrased from \cite{dong_efficient_2021}]\label{prob:sym_qsp}
    Assume we are given a definite parity polynomial $f(x)=\sum_{j=0}^d\alpha_j\cheby_j(x)$, $x\in[-1, 1]$ and initial data $\hat{\Phi}^0=\left(\pi/4, 0, ...,0\right)\in\mathbb{R}^{\tilde{d}}$ for $\tilde{d}=\lceil\frac{d+1}{2}\rceil$. Find $\hat{\Phi}=\left(\phi_0, \phi_1,...\phi_{\tilde{d}-1}\right)$, minimizing the loss function
    \begin{equation}
        L\left(\hat{\Phi}\right)=\frac{1}{\Tilde{d}}\sum_{j=1}^{\Tilde{d}}\left|Re\left[\bra{0}U_{\Phi}(x_j)\ket{0}\right]-f(x_j)\right|^2
    \end{equation}
    over the positive roots of the Chebyshev polynomials $T_{2\Tilde{d}}(x)$, $x_j\in \{\cos\left(\frac{(2j-1)\pi}{4\Tilde{d}}\right)| j=1,...\Tilde{d}\}$.
    $\hat{\Phi}$ is the QSP parameter set  
    \begin{equation}
        {\Phi}=\begin{cases}
            \left(\phi_0, \phi_1,...\phi_{\Tilde{d}-1}, \phi_{\tilde{d}-1},....\phi_{1}, \phi_0, \right) & d \text{ is odd}\\
            \left(\phi_0,...\phi_{\Tilde{d}-2}, \phi_{\tilde{d}-1},\phi_{\Tilde{d}-2}, ....,\phi_0, \right) & d \text{ is even}
        \end{cases}
    \end{equation}
    for $U_{\Phi}$ defined in \cref{eq:Uphi_Gilyen_thrm3}.
    Defining polynomial $f_{\hat{\Phi}}(x)=\sum_{j=0}^d\beta_j\cheby_j(x)=Re\left[\bra{0}U_{\Phi}(x_j)\ket{0}\right]$, the optimized solution will obey $\max_{j=1,...d}\left|\alpha_j-\beta_j\right|\leq 2\sqrt{\varepsilon}$ whenever $L(\hat{\Phi})\leq \varepsilon$.
\end{problem}

 The input function is definite parity and $|P(x)|<1$, and the optimization problem was first solved by quasi-newton methods in \cite{dong_efficient_2021}. One can also use a process called \textit{phase padding}. Essentially, the degree of the polynomial is increased by a factor of $2l$ by adding leading order terms defined by $\phi_{0}=\phi_{n+1+2l}=\pi/4$; for symmetric $\Phi$, this provably preserves the upper left component of the block-encoding \cite{dong_efficient_2021}. Then small phase sets are used to obtain a rough estimate of the function, and then the degree is increased with the phase padded set as an initial guess.\\
 
 What makes these optimization methods powerful is the accompanying results quantifying under what functional conditions the loss function in \cref{prob:sym_qsp} will converge to a global minimum. For example, when $\linfnorm{P(x)}\leq\frac{\sqrt{3}}{20 \pi}\left(\lceil\frac{n+1}{2}\rceil\right)^{-1}$ for gradient descent, and $\linfnorm{P(x)}\propto \frac{1}{n}$ whenever quasi-Newtonian methods are used for the optimization \cite{wang_energy_2022}. \\

An alternative optimization strategy relies on having access to the coefficients of the desired polynomial approximation. Recall that at its simplest, all QSP-processing is a map from a length $n+1$ coefficient vector\footnote{or length $2n+1$ for Laurent-QSP}  to a length $n+1$ set of parameters $\Phi$. Most QSP polynomials are truncations of infinite expansions, and so if the infinite series has coefficient vector $\vec{c}\in\mathbb{R}^{\infty}$, we can define the truncated coefficient vector $c_t$ as another vector in $\mathbb{R}^{\infty}$ with all $c_{i>n}=0$. We can define the coefficient list $f(c_t)=\Phi$ similarly as a vector in $\mathbb{R}$ such that all $\phi_{j>n}=0$.\\

Once $c_t\rightarrow c$ and $c_t\rightarrow \Phi$, $\Phi\rightarrow \Phi_T$ are well-defined mappings on Banach spaces, proving the convergence of the symmetric-QSP loss function can be restated as proving the existence of an inverse mapping $\Phi\rightarrow c$. An elegant result in \cite{dong_infinite_2022} proves this mapping exists and relates the convergence of  $\left|\Phi\right|_1$ to the convergence of $\left|c\right|_1$. The QSP-processing algorithm is a simple fixed-point iteration method. Although a strict condition on the input polynomial coefficients is required to guarantee the protocol succeeds, the numeric work in \cite{dong_infinite_2022} works outperforms the condition. Because the core idea is to define target functions in the  $n\rightarrow\infty$ limit and then truncate the angles resulting in an $\epsilon$-close approximation, this technique is called \textit{infinite-QSP} and the QSP-processing algorithm is called Fixed-Point iteration (FPI). Based on Newton's iterative methods, FPI can succeed with $\mathcal{O}(n\log(n))$ time when the numeric methods of \cite{ni2024fastphasefactorfinding} are used.\\

 The most up-to-date algorithm is presented in \cite{dong2023robustiterativemethodsymmetric}, uses Newton's method for the FPI step, and also writes the QSP circuit as a matrix product state to improve algorithm complexity. It also exploits a homomorphism between $SU(2)$ and a subset of $\mathbb{R}^4$ to write symmetric-QSP circuits with a recurrence on real-valued polynomials, which improves the computational complexity. This QSP-processing is usually called Newton's method\footnote{Multiple QSP-processing processes use a version of Newton-Raphson iteration, including the one discussed at length in \cref{sec:fejer}. Herein, "Newton’s method for finding reduced phase factors", shortened to "Newton's method," will always refer to the method of\cite{dong2023robustiterativemethodsymmetric}} . Dong et al. also introduce the terminology `near-fully coherent' polynomials, which have $\linfnorm{f}=1-\delta$ for a small $\delta\in (0, 1)$. The most up-to-date benchmarks can succeed in nearly fully-coherent polynomials.

\subsubsection{Optimization of Complex Polynomials}\label{sec:brute_opt}
Following \cref{lemma:QSP_Motlagh}, we are given complex polynomial $P$ with scalar coefficient list $\{p\}$, and want to identify $Q$ with coefficients $\{q\}$. The key observation, due to Motlagh and Wiebe in \cite{motlagh_generalized_2023}, is that on the unit circle, the discrete Fourier transforms of $P, Q$ are impulse trains/Dirac combs\footnote{ie, $P(e^{i\theta})=\sum_{k=0}^n p_ke^{ik\theta}$ is the discrete Fourier transform of Dirac Comb operator with period $T=2\pi$ applied to coefficient list $\vec{p}$, where the comb operator is $\text{comb}_{T}\left(f(t)\right)=\sum_{k}={-\infty}^{\infty}f(kT)\cdot \delta\left(t-kT\right)$}. $\left|P(z)\right|^2=P(z)P^*(1/z)$ is a degree $2n$ polynomial with coefficients $\Vec{p}\star\text{reverse}\left(\vec{p}\right)^*$, where $\star$ is the convolution operator. Then \cref{prob:qsp_poly_constraint} can be restated with respect to the coefficients lists $\vec{p}=\{p_j\}, \vec{q}=\{q_j\}$ as
\begin{equation}
    \sum_l\left[\Vec{p}\star\text{reverse}\left(\vec{p}\right)^*+\Vec{q}\star\text{reverse}\left(\vec{q}\right)^*\right]_lz^l=1.
\end{equation}
We can now see that a solution is obtained from the following optimization problem 
\begin{equation}\label{eq:motlaugh_argmin}
    \textrm{argmin}_{\Vec{b}}\left|\left|\Vec{p}\star\text{reverse}\left(\vec{p}\right)^*+\Vec{q}\star\text{reverse}\left(\vec{q}\right)^*-1\right|\right|^2.
\end{equation}
A simple recursion algorithm, similar to the carving step used above, computes $\Phi, \Theta, \lambda$ from the coefficients of $P, Q$. 

This method is extremely fast for real, random polynomial benchmarks \cite{motlagh_generalized_2023}. However, the method does not give a unique solution for at least Hamiltonian Simulation \cite{berntson2024complementarypolynomialsquantumsignal}. Additionally, the convergence of \cref{eq:motlaugh_argmin} over the set of G-QSP relevant polynomials is not as well understood as for the symmetric-QSP problem.

\subsection{Riemann-Hilbert-Weiss Methods}
The Riemann-Hilbert Weiss method is a direct QSP-processing method, but because it relies heavily on the infinite-QSP framework introduced with optimization methods in \cref{sec:symmetric-qsp-processing}, its presentation was deferred.\\

A recent series of papers  \cite{alexis2024quantumsignalprocessingnonlinear, alexis2024infinitequantumsignalprocessing} showed that the nonlinear Fourier transform (NLFT) of desired signal $f(x)$ can be mapped to a set of infinite QSP phase factors when $x\in[0, 1]$ and $f$ is purely real or imaginary. One additional condition is that $f$ satisfy a Szeg{\"o} condition, that is
\begin{equation}
    \int_{0}^{1}\frac{\log\left|1-f(x)^2\right|}{\sqrt{1-x^2}}dx>-\infty.
\end{equation}
This is not difficult for QSP-relevant polynomials to fulfill; any even/odd sum of Chebyshev polynomials will satisfy it. Through standard coordinate conversions, real polynomial $f(x)$ determines $iP(z)$, a purely imaginary Laurent polynomial. The problem of infinite-QSP is then framed as finding a complimentary $Q$, which identifies a NLFT of a sequence of complex numbers. \\ 

The first algorithm based on this approach is known as the Riemann-Hilbert-Weiss algorithm, named after core subroutines. The Weiss algorithm is used to identify the coefficients of a Laurent series with purely imaginary coefficients, this is the completion step. Theorem 5 of \cite{alexis2024infinitequantumsignalprocessing}shows that the resulting pair of Laurent polynomials $(P, Q)$ is a NLFT. Then, Riemann-Hilbert factorization guarantees a unique decomposition determining infinite-QSP parameters $\Phi$, corresponding to the NLFT $(P, Q)$. Subsequently, Ni and Yang improved the computational complexity of the linear equation solver using the half-Cholesky method \cite{ni2024fastphasefactorfinding}; the same work compares their results to the Newton's method and fixed point iteration discussed in \cref{sec:symmetric-qsp-processing}. With the Half-Cholesky method, the Riemann-Hilbert-Weiss strategy can identify angles for symmetric QSP even in the near-fully-coherent regime.\\

This is already a compelling argument for the method. However the half-Cholesky method also has complexity $\Tilde{\mathcal{O}}(d^2+d\eta^{-1}\log(\epsilon^{-1}))$ where $\Tilde{\mathcal{O}}(X)={\mathcal{O}}(X\text{polylog}(X))$, and is numerically stable. It outperforms all other methods on random real Chebyshev series and the Jacobi-Anger expansion of $\cos$, along with the Prony method \cite{Ying2022stablefactorization}, and infinite-QSP optimization strategies (Newton and FPI \cite{dong2023robustiterativemethodsymmetric}), this remains one of the most expansive benchmarking results.
\section{Results}\label{sec:results}
We select two test polynomials: truncated Jacobi-Angler expansions and complex sums of Chebyshev series expansions. All QSP simulations are generated in Python with an AMD Ryzen 7 and 32BG of RAM. Our implementation is available at \url{https://github.com/Skeltonse/QSP-hitchhicker}.

\subsection{Hamiltonian Simulation} 
\begin{figure}[ht!]
    \centering
\begin{tikzpicture}

\definecolor{darkgray176}{RGB}{176,176,176}

\begin{groupplot}[group style={group size=1 by 2, vertical sep=1.5cm}]
\nextgroupplot[
tick align=outside,
tick pos=left,
x grid style={darkgray176},
xlabel={\(\displaystyle \log_{10}(n)\)},
xmin=1.61746896178035, xmax=3.4104918980051,
xtick style={color=black},
y grid style={darkgray176},
ylabel={$-\log_{10}(\epsilon)$},
ymin=-14.1484298677624, ymax=-12.2318512583987,
ytick style={color=black},
width =\linewidth,
height=0.3*\linewidth,
legend style={ at={(0.40,.95)},anchor=north east}
]

\addplot [draw=c1, fill=c1, mark=*, only marks]
table{%
x  y
1.69897000433602 -14.0475909012235
1.95424250943932 -13.6610304719347
2.10380372095596 -13.5344418202352
2.20951501454263 -13.4067203028096
2.29225607135648 -13.3567643724954
2.36172783601759 -13.2617428509138
2.41995574848976 -13.1864866480351
2.47129171105894 -13.1473266549594
2.50105926221775 -13.1154279090263
2.5705429398819 -13.0405577176354
2.62838893005031 -12.9567138361168
2.68033551341456 -12.9273945138992
2.72591163229505 -12.861990429858
2.7664128471124 -12.8292712149297
2.80413943233535 -12.7841747395234
2.83884909073726 -12.7300634661303
2.87040390527903 -12.7091327659342
2.8998205024271 -12.689325473521
2.92737036303902 -12.6441973578015
2.9532763366673 -12.6381381154329
2.97772360528885 -12.6094541425837
3.00086772153123 -12.5796789502558
3.02284061087653 -12.5558227046702
3.04375512696868 -12.5416271516692
3.08242630086077 -12.4978280302979
3.11793383503964 -12.4406791888527
3.15075643986031 -12.4327570138999
3.18127177155946 -12.4089367788141
3.20978301484852 -12.3798547166989
3.23628527744803 -12.342509204692
3.26150077319828 -12.3104407380997
3.28510702956681 -12.2750072857601
3.30749603791321 -12.2675533214932
3.32899085544943 -12.249663870407
};
\addplot [semithick, c2]
table {%
1.69897000433602 -13.9714652185025
1.95424250943932 -13.6990726825107
2.10380372095596 -13.5394810426883
2.20951501454263 -13.426680146519
2.29225607135648 -13.3383900022786
2.36172783601759 -13.2642590646432
2.41995574848976 -13.2021260560867
2.47129171105894 -13.1473472109223
2.50105926221775 -13.1155832779161
2.5705429398819 -13.0414396283237
2.62838893005031 -12.9797141559601
2.68033551341456 -12.9242837382856
2.72591163229505 -12.8756510247958
2.7664128471124 -12.8324335673952
2.80413943233535 -12.7921768220828
2.83884909073726 -12.75513933592
2.87040390527903 -12.7214682754496
2.8998205024271 -12.6900788334328
2.92737036303902 -12.6606813219567
2.9532763366673 -12.6330379459062
2.97772360528885 -12.6069511039441
3.00086772153123 -12.5822548111961
3.02284061087653 -12.5588082945044
3.04375512696868 -12.5364911316157
3.08242630086077 -12.4952264481279
3.11793383503964 -12.4573375760855
3.15075643986031 -12.4223137000854
3.18127177155946 -12.3897518361088
3.20978301484852 -12.3593284661472
3.23628527744803 -12.3310488105842
3.26150077319828 -12.3041422197984
3.28510702956681 -12.2789527931702
3.30749603791321 -12.2550622501626
3.32899085544943 -12.2321258675036
};
\addlegendentry{$-\log_{10}\left(||U_{QSP}-f||_{\infty}\right)$}
\addlegendentry{$1.07\log_{10}(n)-15.78$}

\nextgroupplot[
tick align=outside,
tick pos=left,
x grid style={darkgray176},
xlabel={Polynomial degree \(\displaystyle n\)},
xmin=-54.15, xmax=2237.15,
xtick style={color=black},
y grid style={darkgray176},
ylabel={Completion step time (\(\displaystyle s\))},
ymin=-0.344842249996145, 
ymax=7.83165464998747,
ytick style={color=black},
width =\linewidth,
height=0.3*\linewidth,
]

\addplot [semithick, c1, mark=Mercedes star flipped*, mark size=3, mark options={solid}]
table {%
50 0.0268167000031099
90 0.0451074999873526
127 0.0781520000309683
162 0.0962500000023283
196 0.138493600010406
230 0.152684499975294
263 0.225797299994156
296 0.245087099960074
317 0.189076600014232
372 0.229489799996372
425 0.401269400026649
479 0.498142900003586
532 0.511135499982629
584 0.67473249998875
637 0.779355300008319
690 0.969516500015743
742 1.00211649999255
794 0.763984899967909
846 0.990794399986044
898 1.34772189997602
950 1.16668929997832
1002 1.44615470001008
1054 1.57220600004075
1106 1.72306190000381
1209 2.6592074999935
1312 2.53039059997536
1415 3.71453589998418
1518 4.26073879998876
1621 4.3706745999516
1723 3.64692510000896
1826 4.58401569997659
1928 6.7185204999987
2030 7.45999569998821
2133 5.51745739998296
};
\end{groupplot}

\end{tikzpicture}
    \caption{Solution quality for $P_{HS}$ instances within $\tau=20,...2000$. Left: the log-log plot of solution decomposition error against polynomial degree. Right: completion step times against polynomial degree.}
    \label{fig:hs_times_its}
\end{figure}
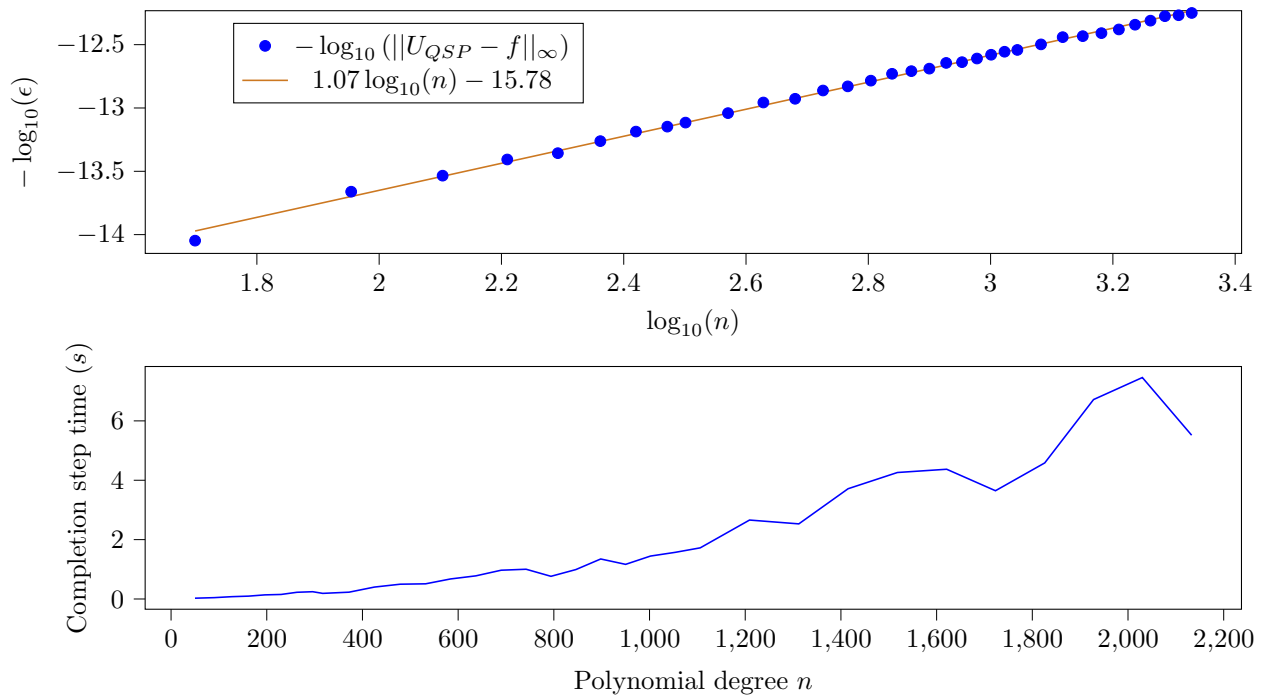

Our target polynomial is found in \cref{sec:hs_poly}. Given precision $\epsilon_{approx}=10^{-14}$, we solve instances with evolution times within $\tau=200, 2000$. \cref{fig:hs_times_its} shows the resulting simulation times, which are competitive with the results of \cite{dong_efficient_2021, Ying2022stablefactorization}. Our truncation degree, required for provably $\epsilon$-bound approximations, leads to longer coefficient lists compared to \cite{dong_efficient_2021}\footnote{For this reason, instances $\tau=4000, 5000$ are outside the space limits of a laptop with machine precision with our truncation, although they are solvable with our method when the truncation proposed in \cite{dong_efficient_2021} is used.}. 

\subsection{Random Polynomials}
\begin{figure}[ht!]
    \centering
\begin{tikzpicture}

\begin{groupplot}[group style={group size=1 by 2, vertical sep=1.5cm}]
\nextgroupplot[
tick align=outside,
tick pos=left,
x grid style={darkgray176},
xlabel={\(\displaystyle \log_{10}(n)\)},
xmin=2.25102999566398, xmax=3.35102999566398,
xtick style={color=black},
y grid style={darkgray176},
ylabel={$-\log_{10}(\epsilon)$},
ymin=-13.7397084801686, ymax=-12.600266305648,
ytick style={color=black},
width =\linewidth,
height=0.3*\linewidth,
 legend style={ at={(0.4,.95)},anchor=north east}
]
\addplot [draw=c1, fill=c1, mark=*, only marks]
table{%
x  y
2.30102999566398 -13.687915654054
2.46834733041216 -13.6660094282538
2.58994960132571 -13.4197943074096
2.68484536164441 -13.2077736244351
2.76192783842053 -13.1342977523822
2.82801506422398 -13.1676492460698
2.88536122003151 -13.0188296402324
2.93601079571521 -13.0286990871128
2.98091193777684 -13.2588906522943
3.02201573981772 -13.1372040374595
3.05956341790127 -13.1364520390185
3.09412159584056 -13.0287036883499
3.12580645813953 -13.0226043484033
3.15563963375978 -12.804775740948
3.18355453361886 -12.9238633392064
3.20978301484852 -12.8428397797826
3.23426412437879 -12.7814603060587
3.25767857486918 -12.6520591317625
3.27989498001164 -12.8888173649074
3.30102999566398 -12.8581243927482
};
\addplot [semithick, c2]
table {%
2.30102999566398 -13.6730564238457
2.46834733041216 -13.5251158496234
2.58994960132571 -13.4175961487342
2.68484536164441 -13.3336901189499
2.76192783842053 -13.265534444023
2.82801506422398 -13.2071006771402
2.88536122003151 -13.1563956905207
2.93601079571521 -13.1116117634875
2.98091193777684 -13.0719105529115
3.02201573981772 -13.0355669183097
3.05956341790127 -13.0023675784758
3.09412159584056 -12.9718115294202
3.12580645813953 -12.9437960417858
3.15563963375978 -12.9174177998733
3.18355453361886 -12.8927356811871
3.20978301484852 -12.8695446796096
3.23426412437879 -12.8478986893304
3.25767857486918 -12.8271958299652
3.27989498001164 -12.8075522721779
3.30102999566398 -12.7888648698445
};

\addlegendentry{\(\displaystyle  -\log_{10}||U_{QSP}-f||_{\infty}\)}
\addlegendentry{$0.88\log_{10}(n)-15.71$}

\nextgroupplot[
tick align=outside,
tick pos=left,
x grid style={darkgray176},
xlabel={Polynomial degree \(\displaystyle n\)},
xmin=110, xmax=2090,
xtick style={color=black},
y grid style={darkgray176},
ylabel={Completion step time (\(\displaystyle s\))},
ymin=-0.156845420025638, ymax=5.43976581999741,
ytick style={color=black},
width =\linewidth,
height=0.3*\linewidth,
]
\addplot [semithick, c1, mark=Mercedes star flipped*, mark size=3, mark options={solid}]
table {%
200 0.0975459999754094
294 0.220207399979699
389 0.314359400013927
484 0.386302799975965
578 0.515765999967698
673 0.851618400018197
768 0.959961299959105
863 1.185240300023
957 1.33885070000542
1052 1.53472709999187
1147 1.83394229999976
1242 2.29449850000674
1336 2.89921479998156
1431 2.92247330001555
1526 3.64898669999093
1621 4.09933020005701
1715 3.23623220005538
1810 4.01620309997816
1905 4.43851869995706
2000 5.18537439999636
};
\end{groupplot}

\end{tikzpicture}
    \caption{Solution quality for random polynomials with degrees within $\left[20, 1000\right]$. Left: The log-log plot of solution error against degree demonstrates that the error in our method scales linearly with the degree. Right: completion step time scales subexponentially with polynomial degrees.}
    \label{fig:random_data}
\end{figure}
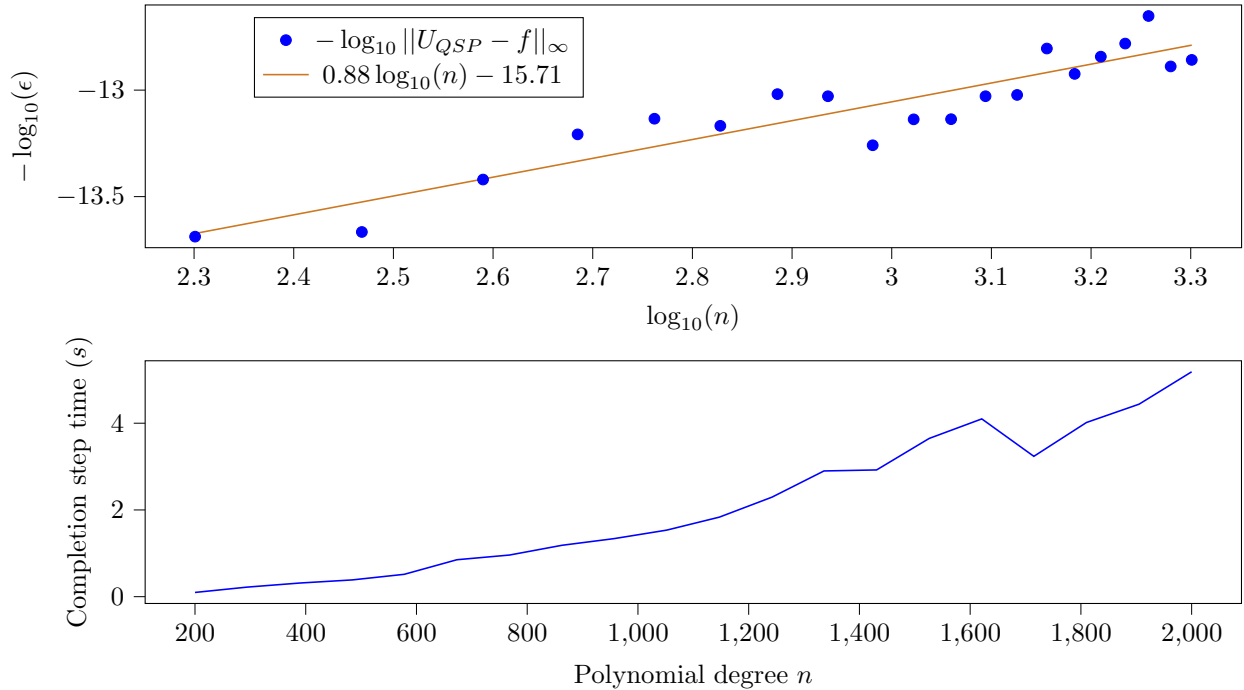

For the randomly generated polynomials described in \cref{sec:random_poly}, we define the polynomial approximation error $\epsilon_{approx}=\mathcal{O}(10^{-12})$ from small imaginary components when building $P_{random}$. We find that the solution error and solution time can be kept below this threshold; see \cref{fig:random_data}. Our instances have a max degree $n=2000$ and obey $\linfnorm{P_{random}}\leq \frac{1}{2}$. However, the barrier to solving QSP-processing is generally the allowed number of nonzero coefficients and the subnormalization, rather than the polynomial degree.

In \cref{sec:wilson}, we showed that the convergence criteria for the Wilson method are quite strict, and in fact, none of our instances obey it. One approach would be to introduce a strict normalization, around $\frac{1}{(n+1)}^2$, which would correspondingly lower the success probability of the QSP protocol. We instead use the solver outside of its known success criteria, and observe that when coefficients obey a decay or sparsity condition, then far greater polynomial degrees are tractable. We find that when subnormalization $1/2$ is imposed, $7$ out of the $180$, i.e., $3.9\%$ of the tested instances fail. 

\subsection{Comparison of G-QSP and QSP-processing methods}
The most relevant comparison to the Wilson method is the BerS{\"u}n method. To our knowledge, this is the only other bounded-error completion step method outside of symmetric-QSP. We also consider the optimization methods originally proposed for G-QSP. We solve and time the completion step using our benchmark polynomials and openly available source code. For each benchmark, we compute the average error over all instances, using
\begin{equation}
   \epsilon=1- \left|{P}\right|^2-\left|{Q}\right|^2.
\end{equation}
This is the loss function of the optimization, and also a reasonable measure of solution quality for the Wilson and BerS{\"u}n methods.
We find that the BerS{\"u}n and Wilson methods are both adequate for most instances. However, the BerS{\"u}n implementation is faster, in fact outperforming our Wilson implementation by $1-2$ orders of magnitude. The optimization method struggles on high precision instances, achieving only $\mathcal{O}(10^{-8})$ precision. The BerS{\"u}n method also copes better with the coherent regime, finding solutions with $l^{\infty}$-norms close to one, whereas in practice, the Wilson method usually requires a subnormalization of $\frac{1}{\sqrt{2}}$ or $\frac{1}{{2}}$.
\pagebreak
\begin{figure}[ht]
     \centering
     \begin{subfigure}[b]{\textwidth}
         \centering
\begin{tikzpicture}

\begin{axis}[
tick align=outside,
tick pos=left,
x grid style={darkgray176},
xlabel={degree},
xmin=-54.15, xmax=2237.15,
xtick style={color=black},
y grid style={darkgray176},
ylabel={completion step time},
ymin=-1.31715607286897, ymax=30.0093632055155,
ytick style={color=black},
width =\linewidth,
height=0.3*\linewidth,
 legend style={ at={(0.4,.99)},anchor=north east}
]
\addplot [semithick, dashdotted,c2]
table {%
50 0.0201639999868348
90 0.0441975999856368
127 0.0593283000052907
162 0.100172000005841
196 0.140019499987829
230 0.16426320001483
263 0.185503999993671
296 0.238589199958369
317 0.215400500048418
372 0.28982070001075
425 0.296132199990097
479 0.35194149997551
532 0.413252199999988
584 0.452836300013587
637 0.52965400001267
690 0.789927100006025
742 0.961016800021753
794 1.35695500002475
846 0.961697499966249
898 1.60972810001113
950 1.60962930001551
1002 1.90782710001804
1054 2.11903230001917
1106 2.02130160003435
1209 2.32686749997083
1312 2.86748110002372
1415 2.41642909997609
1518 4.05845569999656
1621 3.42985610000324
1723 5.21880690002581
1826 4.45893299998716
1928 5.45912770001451
2030 5.98399560002144
2133 5.7161709999782
};
\addplot [semithick,  c1]
table {%
50 0.0106152000371367
90 0.0287768000271171
127 0.0176220000139438
162 0.0647745000314899
196 0.0537775999982841
230 0.110987499996554
263 0.0323166999733075
296 0.00999769999179989
317 0.0643335999920964
372 0.137869699974544
425 0.0160308000049554
479 0.0418557999655604
532 0.0967696000006981
584 0.0982631000224501
637 0.115189900039695
690 0.1131299000117
742 0.0290699999895878
794 0.18425190000562
846 0.252951200003736
898 0.17186290002428
950 0.059086200024467
1002 0.221644999983255
1054 0.0423790999921039
1106 0.187690799997654
1209 0.28760440001497
1312 0.050224700011313
1415 0.29459320002934
1518 0.0949203999480233
1621 0.0882091000094078
1723 0.107933600025717
1826 0.352744599978905
1928 0.143412099976558
2030 0.32931160001317
2133 0.334527600032743
};
\addplot [semithick, dashed, c3]
table {%
50 2.50334405899048
90 1.68711471557617
127 1.93771743774414
162 2.61833453178406
196 1.54297661781311
230 3.40131735801697
263 2.98972201347351
296 4.1943154335022
317 2.27608036994934
372 4.64711308479309
425 4.65817785263062
479 8.40888595581055
532 4.31180906295776
584 3.07734489440918
637 5.76583361625671
690 5.48669195175171
742 5.46858763694763
794 14.8660428524017
846 20.7497217655182
898 12.0888991355896
950 18.3728175163269
1002 9.32326722145081
1054 11.2659339904785
1106 11.5274333953857
1209 28.2915439605713
1312 14.2428822517395
1415 11.8944928646088
1518 16.4095597267151
1621 13.0228984355927
1723 20.8087167739868
1826 9.29979038238525
1928 14.0914971828461
2030 20.866551399231
2133 28.5808219909668
};
\addlegendentry{Wilson: $\braket{\epsilon}=1.1\pm 0.9 \cdot 10^{-13}$}
\addlegendentry{FFT: $\braket{\epsilon}=2\pm 2 \cdot 10^{-13}$}
\addlegendentry{Opt: $\braket{\epsilon}=2.8\pm 1.0 \cdot 10^{-8}$}

\end{axis}

\end{tikzpicture}
     \end{subfigure}
     \hfill
     \begin{subfigure}[b]{\textwidth}
         \centering
\begin{tikzpicture}

\begin{axis}[
tick align=outside,
tick pos=left,
x grid style={darkgray176},
xlabel={degree},
xmin=-29, xmax=1049,
xtick style={color=black},
y grid style={darkgray176},
ylabel={completion step time},
ymin=-1.38114218733972, ymax=29.0407633350114,
ytick style={color=black},
width =\linewidth,
height=0.3*\linewidth,
 legend style={ at={(0.4,.95)},anchor=north east}
]
\addplot [semithick, dashdotted, c2]
table {%
20 0.00986120000015944
50 0.0210655999835581
80 0.038229700003285
110 0.067744399944786
140 0.0684445999795571
170 0.107575499976519
200 0.127241800015327
230 0.152146399952471
250 0.16398820001632
300 0.258350900025107
350 0.31149860000005
400 0.340945000003558
450 0.426044600026216
500 0.636426199984271
550 0.679656699998304
600 0.861101999995299
650 0.697560200002044
700 1.02837429998908
750 1.17843199998606
800 0.916983900009654
850 1.43822800001362
900 1.35843900003238
950 1.73799199995119
1000 1.8162078000023
};
\addplot [semithick, c1]
table {%
20 0.00370159995509312
50 0.00167170003987849
80 0.00173819996416569
110 0.00183540000580251
140 0.0100036999792792
170 0.00327560002915561
200 0.0311178999836557
230 0.0315741000231355
250 0.0689424999873154
300 0.0392167000100017
350 0.0378359000314958
400 0.0657117000082508
450 0.00835449999431148
500 0.0396614000201225
550 0.0403521999833174
600 0.0840870999963954
650 0.0460149999707937
700 0.0991714999545366
750 0.0135590000427328
800 0.100449099962134
850 0.043950000021141
900 0.0137614000122994
950 0.144488800026011
1000 0.108680500008631
};
\addplot [semithick, dashed, c3]
table {%
20 1.91243934631348
50 3.10534381866455
80 4.16675686836243
110 1.03181529045105
140 4.41621065139771
170 3.73993349075317
200 3.55953216552734
230 4.55540561676025
250 4.51656198501587
300 5.91860246658325
350 6.73445105552673
400 7.60888361930847
450 12.3357722759247
500 9.03038096427917
550 13.8763217926025
600 8.1550714969635
650 8.15836215019226
700 12.4474332332611
750 21.1026794910431
800 17.8211233615875
850 20.106508731842
900 19.2477931976318
950 27.6579494476318
1000 21.9341776371002
};
\addlegendentry{Wilson: $\braket{\epsilon}=5\pm 3 \cdot 10^{-14}$}
\addlegendentry{FFT: $\braket{\epsilon}=3\pm 3 \cdot 10^{-14}$}
\addlegendentry{Opt: $\braket{\epsilon}=2\pm 2 \cdot 10^{-7}$}

\end{axis}

\end{tikzpicture}
     \end{subfigure}
        \caption{Completion step times for optimization, BerS{\"u}n, and NR methods. Above: Jacobi-Anger expansion. Below: random Chebyshev series}
        \label{fig:method-comp}
\end{figure}
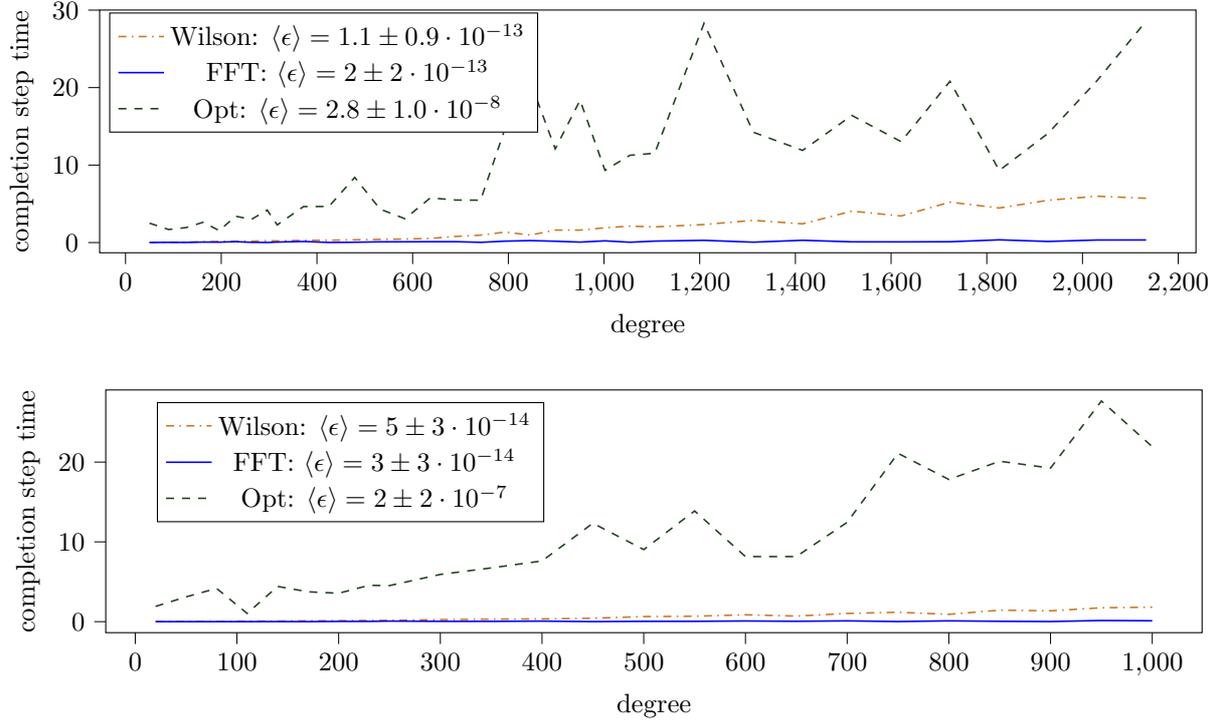

\subsection{Metasolver for QSP and G-QSP processing}
We summarize which QSP-processing methods are currently available to users working with polynomials fitting the conditions of \cref{lemma:QSP_Haah}. Readers can find a more comprehensive discussion of QSP-processing benchmarks for real polynomials in \cite{ni2024fastphasefactorfinding}.

\begin{figure*}[ht!]
    \centering
\scalebox{0.7}{
\begin{tikzpicture}[node distance=2cm]
\node (in1) [io] {target function $f\in \mathbb{C}$};
\node (pro33) [process, below of=in1, xshift=-15]{Require numeric stability?};
\node (pro13) [process, below of=pro33]{Low precision acceptable?};
\node (pro43) [process, below of=pro13]{reciprocally conditions on Laurent form and $\linfnorm{f}\leq \frac{1}{2}$?};
\node (dec1) [decision, right of=pro33, xshift=8cm] {No methods, break into real constituent target functions};
\node (dec2) [decision,  below of=dec1] {optimization \cite{motlagh_generalized_2023}, complex Prony \cite{yamamoto2024robust}};
\node (dec3) [decision, below of=dec2] {Wilson Method};
\node (dec4) [decision, below of=pro43 ] {BerS{\"u}n Method};

\draw [arrow] (in1) --  (pro33);
\draw [arrow] (pro33) -- node {yes} (dec1);
\draw [arrow] (pro33) -- node {no} (pro13);
\draw [arrow] (pro13) -- node {yes} (dec2);
\draw [arrow] (pro13) -- node {no} (pro43);
\draw [arrow] (pro43) -- node {yes} (dec3);
\draw [arrow] (pro43) -- node {no} (dec4);
\end{tikzpicture}
}
    \caption{A decision tree for possible QSP-processing strategies}
    \label{fig:towel_qsp}
\end{figure*}
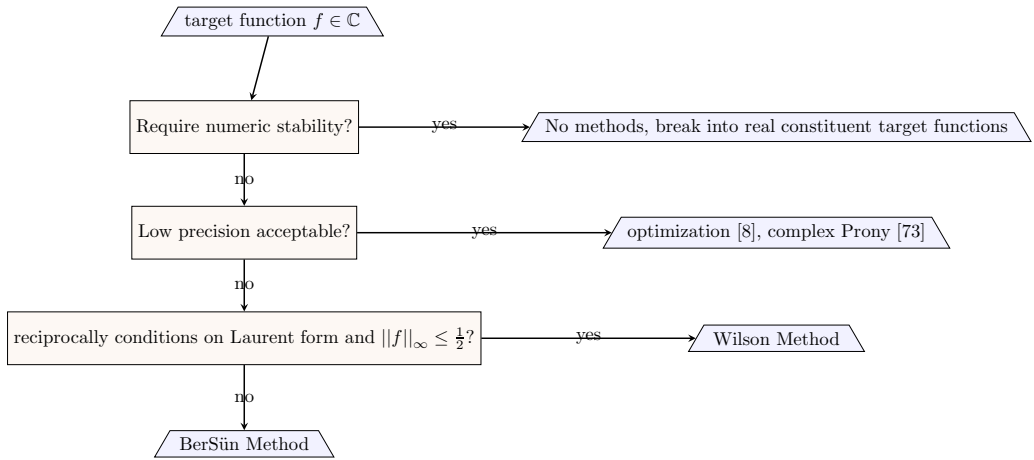

The flow-chart \cref{fig:towel_qsp} summarizes currently available methods for QSP-processing under different user conditions. The strongest strategy for QSP-processing with polynomials of the form \cref{lemma:QSP_Haah} is currently the BerS{\"u}n method in \cite{berntson2024complementarypolynomialsquantumsignal}. Truly numerically stable methods for QSP require users to work with symmetric-QSP, which for complex polynomials forces an increase circuit length and decreased success probability.

\section{Discussion}
QSP-processing has become a progressively more complicated topic within the past five years. Initially, it was thought unpractical even for Hamiltonian simulation \cite{childs2018towards}. Improvements were introduced to solve some moderately challenging instances \cite{martyn_efficient_2023, chao2020finding, Haah2019product}, and both optimization and direct methods have made larger instances tractable for symmetric-QSP \cite{dong_efficient_2021, dong_infinite_2022, Ying2022stablefactorization, alexis2024infinitequantumsignalprocessing}. \\

These successes have shifted attention and effort towards symmetric-QSP circuits, which, for complex polynomials, slightly scale down the success probability and increase the circuit depth. In our review of QSP-processing, we have outlined a few key choices for the user: what conditions the polynomial obeys, what precision and subnormalization are acceptable, and how simple a circuit is desired. Working with symmetric-QSP provides a greater set of pre-processing strategies, but for complex functions, it introduces a subnormalization that carries through to the success probability. One central motivating question of this work was whether bounded precision methods could be made competitive for complex input polynomials and, if so, which strategy is optimal. \\

Our discussion of the Wilson method for QSP and \cref{thrm:laurent_qsp_with_processing} is sufficient to show that QSP can be used to prepare a circuit embedding a complex polynomial within bounded error. We have modified the algorithm developed by Haah in \cite{Haah2019product}, which characterizes how errors in the polynomial approximation, coefficient encoding, and completion step jointly propagate through the decomposition step. However, using the Wilson method for the completion step has some disadvantages \textemdash reciprocity restricted polynomials and $\mathcal{O}(n^3)$ floating point operations, as well as an implicit sub-normalization requirement. Recent work \cite{berntson2024complementarypolynomialsquantumsignal} has $\mathcal{O}(n\log n)$ time and succeeds well for all of our benchmark cases. Optimization techniques, without using the cost landscape restrictions accessible to symmetric-QSP, also work in  $\mathcal{O}(n\log n)$ time but struggle with high precision instances. Altogether, for arbitrary complex polynomials whose norm is bounded on $U(1)$, the most plausible approach is likely to use \cite{berntson2024complementarypolynomialsquantumsignal} for the completion step.\\

The decomposition error developed in this work, which ensures that the output rotations are exactly unitary, is actually agnostic to the completion method used. Thus, optimization or BerS{\"u}n methods can be employed to bound the error in a Laurent-QSP decomposition step. Alternately, one could use any completion step discussed herein and explicitly bound how the completion step error propagates through the G-QSP decomposition, where a similar result to our \cref{lemma:decomp_error} should hold.\\

 It remains to be seen whether QSP and QSVT-based algorithms can be made competitive enough for the template to be widely adopted; in the meantime, resource analyses of QSP and QSVT algorithms are becoming more common. Reliable and fast QSP-processing methods for both real and complex polynomials are necessary so that such work is not constrained by pre-processing in the most problem-efficient templates. In particular, this means that users do not need to resort to the added circuit complexity and subnormalized success probability of symmetric-QSP in order to get reliable solutions with bounded error. We have shown that several methods of QSP processing are successful for a class of highly relevant complex polynomials up to polynomial degrees and precisions that push the limits of floating point precision.

\section*{Acknowledgements}
The author thanks Tobias Osborne for detailed discussions on the project and John Martyn and Ugn{\.e} Liaubait{\.e} for providing helpful feedback on an early version of the manuscript.  This work was supported by the BMWi project ProvideQ.\\

I also acknowledge the support of the Natural Sciences and Engineering Research Council of Canada (NSERC), PGS D - 587455 - 2024. Cette recherche a été financée par le Conseil de recherches en sciences naturelles et en génie du Canada (CRSNG), PGS D - 587455 - 2024.

\bibliography{qsvtbib}
\appendix
\section{QSP Polynomial Basics}
\label{sec:laurent_poly_review}
We will often require the $l_{\infty}$ norm, denoted $||f||_{\infty}$, and computed as $||f||=\max_{x\in\mathcal{D}}|f(x)|$. Chebyshev polynomials of degree $n$ are defined as follows: Chebyshev polynomials of the first kind are 
\begin{equation}
    \cheby_l(\cos\theta):=\cos\left(l\theta\right),
\end{equation}
and the second kind,
\begin{equation}
    \chebyt_l(\cos\theta):=\frac{\sin((l+1)\theta)}{\sin\theta}.
\end{equation}
For convenience, we often work in $x=\cos\theta$, $\sqrt{1-x^2}=\sin\theta$. Then, we can construct combinations of Chebyshev terms, such as
\begin{equation}\label{eq:cheby_poly}
    P(x)=\sum_{l=0}^n\mathcal{A}_l\cheby_l(x)+\sqrt{1-x^2}\sum_{l=0}^n\mathcal{B}_l\chebyt_l(x).
\end{equation}
The square root is included so that in the original coordinates, 
\begin{equation}\label{eq:cs_poly}
    P(\theta)=\sum_{l=0}^n\mathcal{A}_l\cos(l\theta)+\sin\theta\sum_{l=0}^n\mathcal{B}_l\sin((l+1)\theta).
\end{equation}
Of course, these combinations of Chebyshev polynomials are polynomials $P: x\in[-1, 1]\rightarrow \mathbb{C}$ with coefficients in $\mathbb{C}$. We will often restrict ourselves to $P: x\in[-1, 1]\rightarrow \mathbb{R}$, where standard definitions of even, odd, or bounded functions apply. Furthermore, it is often useful to distinguish between the cosine term, which is always even and is real when the coefficients $\mathcal{A}_l$ are real, and the sinusoidal term, which is always odd and is purely imaginary when $\mathcal{B}_l$ are real. Interested readers can refer to \cite{tang_cs_2023} for an overview of why Chebyshev combinations arise as a natural framework for developing polynomial approximations in QSVT.

We will also extensively use Laurent polynomials of degree $n\geq 0$, 
\begin{equation}
    \mathcal{F}(z)=\sum_{k=-n}^nc_kz^k,
\end{equation}
where coefficients $c_k$  can be matrix or scalar-valued. Such polynomials are defined for $\mathbb{C}-\{0\}$, but we will be primarily concerned with their behavior only on $z=e^{i\theta/2}\in U(1)$.

Following \cite{Haah2019product}, we introduce the following terminology
\begin{remark}
    Laurent polynomial $\mathcal{F}(z)$ defined on $z\in \mathbb{C}-\{0\}$ is said to be reciprocal if $\mathcal{F}(z)=\mathcal{F}(z^{-1})$ and anti-reciprocal if $\mathcal{F}(z)=-\mathcal{F}(z^{-1})$. Furthermore, $F(z)$ is
\begin{itemize}
    \item real if all $C_i\in\mathbb{R}, \mathbb{M}(\mathbb{R})$ 
    \item real-on-circle if $F(z)\in\mathbb{R}$ for all $z\in U(1)$
    \item pure if it is real-on-circle and reciprocal or real-on-circle and $i\mathcal{F}(z)$ is anti-reciprocal
    \item Within the following three properties, any two imply the third: reciprocal, real, and real-on-circle
\end{itemize}
\end{remark}
An important class of Laurent polynomials for QSP is 
\begin{equation}\label{eq:laur_poly_sdecomp}
    \mathcal{F}(z)=\mathcal{A}(z)+i\mathcal{B}(z),
\end{equation}
defined with real-on-circle Laurent polynomials, $\mathcal{A}(z), \mathcal{B}(z)$ such that $\mathcal{A}(z)=\sum_{l=-n}^n{a}_lz^l$ is reciprocal, and $\mathcal{B}(z)=\sum_{l=-n}^n{b}_lz^l$ is anti-reciprocal.

$\mathcal{A, B}$ are easily converted into degree $n$ Chebyshev polynomials with coordinate change $z=e^{i\theta}\rightarrow x=\cos\theta$. 
\begin{remark}
    Consider real-on-circle Laurent polynomials, $\mathcal{A}(z), \mathcal{B}(z)$ such that $\mathcal{A}(z)=\sum_{l=-2n}^{2n}{a}_lz^l$ is reciprocal, and $\mathcal{B}(z)=\sum_{l=-2n-1}^{2n}{b}_lz^l$ is anti-reciprocal.
    Define the following Chebyshev polynomials
    \begin{align}
        f_+(x)=\sum_{l=0}^{n}f_{+,2l}\cheby_{2l}(x)\\
        f_-(x)=\sqrt{1-x^2}\sum_{l=1}^nf_{-,2l+1}\chebyt_{2l}(x)
    \end{align}
    where $\mathcal{A}_l=f_{+,l}/2$ and $\mathcal{B}_l=-if_{-,l}/2$, $x\in[-1, 1]$ both functions are bounded, $\mathcal{A}(x)$ is even, and $\mathcal{B}(x)$ is odd.
    Then on domain $z=e^{i\theta}\in U(1)$, corresponding to $x=\cos\theta \in [-1, 1]$,  $\mathcal{A}(z)=f_+(x)$ and $\mathcal{B}(z)=f_-(x)$.
\end{remark}
From this, it is obvious that equations of the form \cref{eq:cheby_poly} can be written as Laurent polynomials of the form \cref{eq:laur_poly_sdecomp}. The conditions imposed on polynomials in QSP often guarantee that this transformation is possible; it is always true beginning from Chebyshev polynomials. However, some care should be taken because the most general version of Laurent QSP ($\mathcal{A}, \mathcal{B}$ could have the same parity) will not always obey this form. Finally, note that all Fourier expansions can be put in either form.

\section{Target Polynomials}\label{sec:poly_approx}
\subsection{Random Complex Polynomials}\label{sec:random_poly}
We generate polynomials of the following types:
\begin{align}
    P_{Rand}(x)&=\sum_{j=0}^{n}a_{j}\cheby_{j}(x)+i\sum_{j=1}^{n}b_{j}\sin(\arccos(x)) U_{j}(x)\\
    &=\sum_{i=0}^{n}a_{j}\cos(j\theta)+i\sum_{i=0}^{n}b_{j}\sin(j\theta)
\end{align}
where $\cheby_j(x), \chebyt_j(x)$ are respectively Chebyshev polynomials of the first and second kind. Such polynomials are popular representations of QSP functions and, hence, represent a good benchmark. $\{a_j\}, \{b_j\}$ are constructed by sampling over uniform distribution $[0, 1)$. To make the problem computationally tractable, we restrict the maximum number of nonzero coefficients over both polynomials to be $nz=\max(5, n/15)$ and stipulate that the coefficients decay at a rate of $\frac{3}{2}$. Finally, we normalize $\linfnorm{P_{rand}}\leq \frac{1}{2}$, and then it can easily be shown to fulfill the conditions for any QSP procedure. 

\subsection{Hamiltonian Simulation}\label{sec:hs_poly}
Hamiltonian simulation is one of the most studied QSP/QSVT applications and has become a standard QSP-processing benchmark. The target function for simulation time $\tau$ is $e^{i\tau x}$,  and the Jacobi-Anger expansion is used to construct a target polynomial. The expansion is
\begin{align}
e^{i\tau x}&=\cos(\tau{x})+i\sin(\tau{x})\\
    \cos(\tau x)&=J_0(\tau)+2\sum_k(-1)^kJ_{2k}(\tau)\cheby_{2k}(x)\\
    \sin(\tau x)&=2\sum_k(-1)^kJ_{2k+1}(\tau)\cheby_{2k+1}(x),\\
\end{align} 
where $J_{j}(x)$ is the $j$th order Bessel function. We cannot exactly solve for the truncation degree $R$, which bounds precision $\epsilon$ - it would require a solution related to the W-lambeth function. However from \cite{gilyen_quantum_2019}, we can  obtain the following upper bounds

\begin{equation}
        \Tilde{r}(\tau, \epsilon)=\begin{cases}
            \lceil e\tau\rceil, & \tau\geq \frac{\ln(1/\epsilon)}{e}\\
             \lceil \frac{4\ln(1/\epsilon)}{\ln\left(e+\frac{1}{\tau}\ln(1/\epsilon)\right)}\rceil, & \tau\leq \frac{\ln(1/\epsilon)}{e}\\
        \end{cases}
    \end{equation}
    Moreover, for all $q\in\mathbb{R}_+$,
    \begin{equation}
        r(\tau, \epsilon)<e^q\tau+\frac{\ln(1/\epsilon)}{q}.
    \end{equation}

We use
\begin{equation}
        \Tilde{r}(\tau, \epsilon)=\begin{cases}
           e\tau+{\ln(1/\epsilon)} & \tau\geq \frac{\ln(1/\epsilon)}{e}\\
             \lceil \frac{4\ln(1/\epsilon)}{\ln\left(e+\frac{1}{\tau}\ln(1/\epsilon)\right)}\rceil & \tau\leq \frac{\ln(1/\epsilon)}{e}\\
        \end{cases}
    \end{equation}
and then define (in the $z$-basis)
\begin{align}
    P_{HS}(z,\tau, \epsilon)&=J_0(\tau)/2+\sum_{k=-\Tilde{r}}^{\Tilde{r}}(-1)^kJ_{2k}(\tau)z^{2k}\nonumber\\
    &+i\sum_k^{\Tilde{r}}(-1)^kJ_{2k+1}(\tau)z^{2k+1}
\end{align}
A subnormalization to make the QSP-processing tractable is common - we use $\frac{1}{\sqrt{2}}$, which is a slight improvement to benchmarks \cite{Ying2022stablefactorization, dong_infinite_2022}. However, the BerS{\"u}n method is capable of solving polynomials with norms much closer to $1$.\\

For use in symmetric-QSP/QSVT, one would compute solution sets $\Phi_{\cos}$, $\Phi_{\sin}$ corresponding to purely even,odd complex completions of the $\cos, \sin$ expansions, and use LCU to add the two symmetric-QSP circuits together on the circuit. This LCU step introduces a subnormalization of $\frac{1}{2}$, which can be handled using amplitude amplification, for see example \cite{gilyen_quantum_2019}. Because $\cos, \sin$ fits the criteria of target polynomials for \cref{lemma:QSP_Haah},  this is an application where we would like Laurent-QSP to offer a speedup. Note that the more efficient decomposition used in G-QSP in \cite{motlagh_generalized_2023, berry2024doubling} can be used to obtain a circuit with query complexity $n+2$ with respect to oracles $CU, CU^{\dag}$ and $U^{\dag}CU^2$ \cite{berry2024doubling}, however, since this is incompatabile with the reciprocity conditions on Laurent-QSP, we disregard it herein.

\section{Wilson method convergence}\label{sec:wilson-convergence}
\WILSONCONV*
\begin{proof}
We will derive conditions for which the error in each coefficient of $\gamma$ can converge at rate $1-\delta$. This is actually a stronger criterion than used in our algorithm, where we only bound the error in the function.
Denote $|e_a^{(t)}|$, the absolute maximum in any coefficient of error polynomial $e^{(t)}=\Tilde{\gamma}-\gamma^{(t)}$ at the $t$-th step of the iteration, and then we begin with a relation from \cite{wilson_factorization_1969},
\begin{equation}
    |e_a^{(t+1)}|\leq \frac{\rho |e_a^{(t)}|^2}{1-2\rho|e_a^{(t)}|}
\end{equation}

where $\rho=(n+1)\left|\left|(T^*)^{-1}\right|\right|_r$ and $\left|\left|\cdot \right|\right|_r$ is the the matrix row sum norm. To get better than linear convergence, ie $\frac{|e_a^{(t+1)}|}{|e_a^{(t)}|}\leq 1-\delta$ for some $\delta\in(0, 1]$, it is enough to require 
\begin{equation}
    (3-2\delta)\rho |e_a^{(t)}|\leq  1-\delta.\label{eq:wilsonconvrestruct}
\end{equation}
That is, we want to show that  $(3-2\delta)\rho |e_a^{(0)}|\leq  1-\delta$, beginning from $|e_a^{(0)}|=\max \{|\gamma^{(*)}_{0}-\gamma^{(0)}_{0}|, \left(\gamma^{(*)}_{j>0}\right)\}$. Define $B=\frac{1-\delta}{3-2\delta}$ so that the desired bound is $\rho |e_a^{(0)}|\leq B$ or  $\frac{|e_a^{(0)}|}{B}\leq \frac{1}{\rho}$.  Then recall that $\left|\left|(T^*)^{-1}\right|\right|_r\geq \frac{(n+1)}{\left|\left|(T^*)\right|\right|_r}$, and so using the multiplicative property of matrix norms
\begin{equation}
    \frac{1}{\rho}\leq \frac{1}{(n+1)}\frac{1}{\left|\left|(T^*)^{-1}\right|\right|_r}\leq \frac{1}{(n+1)}\frac{\left|\left|(T^*)\right|\right|_r}{n+1}.\label{eq:wilsonconvrhobound}
\end{equation}
Plugging \cref{eq:wilsonconvrhobound} in \cref{eq:wilsonconvrestruct} ,
\begin{align}
  {|e_a^{(0)}|}&\leq \frac{B\left|\left|(T^*)\right|\right|_r}{(n+1)^2}.
\end{align}
Select initial guess $\gamma^{(0)}_0=\gamma^{(*)}_0+\gamma^{(*)}_J$ where $J$ denotes the index of the coefficient of  $\gamma^{(*)}$ with the largest absolute value.  Then, the condition becomes
\begin{align}
  {|\gamma^{(*)}_J|}&\leq \frac{B\left|\left|(T^*)\right|\right|_r}{(n+1)^2},
\end{align}
or,
\begin{align}
  \frac{|\gamma^{(*)}_J|}{\left|\left|(T^*)\right|\right|_r}&\leq \frac{B}{(n+1)^2}.
\end{align}

\end{proof}
In \cref{lemma:wilson-convergence}, we have not required quadratic convergence, which is usually the desired rate for Newton-Raphson iteration (and is, in fact, the expected performance of the method \cite{wilson_factorization_1969, goodman_spectral_1997}). The method's reliance on a linear solver means that, even with a quadratic convergence rate, its convergence is usually sub-optimal compared to \cite{bastidas2024complexificationquantumsignalprocessing} even with quadratic convergence.

\section{Laurent QSP Proof}\label{sec:Haah_proof}
Our proof of \cref{lemma:QSP_Haah} will closely follow \cite{Haah2019product}; we repeat it here to make the subsequent error analysis more comprehensible and because replacing the root-finding completion step with the Fejer method changes the error analysis somewhat. Unlike  \cite{Haah2019product}, we do not explicitly distinguish between float and arbitrary precision arithmetic, and we introduce an additional approximation in the decomposition step to ensure each $P$ is exactly a projector. Both proofs rely on having two real-on-circle, (anti-)reciprocal polynomials $\mathcal{A, B}$ as inputs, but is completely agnostic towards whether $\mathcal{A, B}$ are both (anti-)reciprocal, or one is reciprocal and the other anti-reciprocal. 

\subsection{Idealized Circuit Derivation}\label{sec:haah_proof_noerror}
First we prove that \cref{lemma:QSP_Haah} is correct when an exact solution to the completion step is available. Lemma 3 is essentially Theorem 2 and Lemma 4 from \cite{Haah2019product}, stated as one lemma for convenient comparison to the other QSP lemmas.
\QSPHAAH*
\subsubsection{Step 1: find c,d}\label{sec:haah_proof_step1}
Define 
$$\mathcal{F}(z)=1-\mathcal{A}(z)^2-\mathcal{B}(z)^2,$$ a  reciprocal, real-on circle (and hence, real), degree $n'\leq 2n$ Laurent polynomial. $\mathcal{F}$ has $2n'$ roots, which all have even multiplicity\footnote{This follows from the reciprocity of $\mathcal{F}$, and is the reason for that condition in the lemma.} and none of which are on the unit circle. Thus roots come in reciprocal pairs $(r, r^{-1})$, with one root inside the unit circle and one outside of it. Construct the list of all roots $|r|>1$, ${D}$. Crucially, ${D}$ is closed under complex conjugation\footnote{Note that we can get a different construction using a different list of roots ${D}$ - in fact, Haah uses all the roots inside of the unit disk.}. Define the polynomial constructed from all roots outside of the unit circle, 
$$\gamma(z)=z^{-\lfloor n'/2\rfloor}\prod_{r\in {D}}(z-r).$$
Prefactor $z^{-\lfloor n'/2\rfloor}$ is included to balance the exponents because we define $\gamma$ as a Laurent polynomial of degree $n'/2$, rather than a polynomial of degree $n'$. 
$\gamma(1/z)$ contains all the roots of $\mathcal{F}$ inside of the unit circle, and so 
\begin{equation}
\label{mathcalF_gamma_def}
    \mathcal{F}(z)= \gamma(z)\gamma(1/z)
\end{equation}
On the unit circle, $\mathcal{F}=\left|\gamma(z)\right|^2\geq 0$, and recall
Re-arrange 
\begin{align}
    \mathcal{F}&=\left(\frac{\gamma(z)+\gamma(1/z)}{2}\right)^2+\left(\frac{\gamma(z)+\gamma(1/z)}{2i}\right)^2
\end{align}
and define
\begin{align}
    \mathcal{C}(z)&=\frac{\gamma(z)+\gamma(1/z)}{2}\\
    \mathcal{D}(z) &=\frac{\gamma(z)-\gamma(1/z)}{2i}.
\end{align}
where by construction $\mathcal{C, D}$ are degree $\lceil n'/2\rceil$ Laurent polynomials, real-on-circle , and respectively reciprocal and anti-reciprocal.

\subsubsection{Step 2: Build F(z)}\label{sec:haah_proof_step2}
We have set $\mathcal{A, B, C}, \mathcal{D}$ of real-on-circle polynomials such that we can define a matrix coefficient Laurent polynomial
\begin{align}
    F(z)&=\mathcal{A}(z)I+i\mathcal{B}(z)\sigma_x+i\mathcal{C}(z)\sigma_y+i{\mathcal{D}}(z)\sigma_z\\
    &=\sum_{j=-n}^n\tilde{C}_jz^j
\end{align}
and guarantee that $F(z)\left(F(z)\right)^{\dag}=I.$ Furthermore, $F$ has the largest degree of $\mathcal{A, B, C, D}$.

\subsubsection{Step 3: decompose F(z)}\label{sec:haah_proof_step3}
$\text{det}\left(F(z)\right)=1$ $\forall z$, and so we must have
\begin{align*}
    1&=\text{det}(\tilde{C}_{-n})z^{-2n}+\text{higher order terms},\\
     1&= \text{lower order terms} + \text{det}(\tilde{C}_{n})z^{2n}.
\end{align*}
For this to hold in general we require $\text{det}(\tilde{C}_{n}), \text{det}(\tilde{C}_{-n})=0$.
Performing the analogous expansion on $F(z)\left(F(z)\right)^{\dag}=\left(F(z)\right)^{\dag}F(z)=I$, the leading order terms are $z^{-2n}\tilde{C}_{-n}\tilde{C}_{n}^{\dag}, z^{2n}\tilde{C}_{n}\tilde{C}_{-n}^{\dag}$ which introduces constraints
\begin{equation}
    \tilde{C}_{-n}^{\dag}\tilde{C}_{n}=0 \quad \tilde{C}_{n}\tilde{C}_{-n}^{\dag}=0
\end{equation}
However, since $F(z)$ is a degree $n$ polynomial, at least one of $\tilde{C}_n, \tilde{C}_{-n}\neq 0$. Say $\tilde{C}_n\neq 0$, and then because $\tilde{C}_n$ is a rank-1 2x2 matrix, there exists a unique choice of rank-1 projector $P$ such that $\tilde{C}_nP=0$. Both $\tilde{C}_{-n}^{\dag}, P$ annihilate $\tilde{C}_n$, and so simple computation $\tilde{C}_n\tilde{C}_{-n}^{\dag}\tilde{C}_{-n}=0$ implies that $P\propto \tilde{C}_{-n}^{\dag}\tilde{C}_{-n}$ for $\tilde{C}_{-n}\neq 0$\footnote{See Lemma 2 in \cite{Haah2019product} for more careful justification of $\tilde{C}_{-n}=0$}. Thus we define normalized $P=\frac{\tilde{C}_{-n}^{\dag}\tilde{C}_{-n}}{\text{Tr}\left(\tilde{C}_{-n}^{\dag}\tilde{C}_{-n}\right)
}$, and there exists a unique orthonormal projector $Q= I-P$. From the constraints we must also have $\tilde{C}_n^{\dag}\tilde{C}_nP=0$ so then $Q=\frac{\tilde{C}_{n}^{\dag}\tilde{C}_{n}}{\text{Tr}\left(\tilde{C}_{n}^{\dag}\tilde{C}_{n}\right)
}$ and $\tilde{C}_{-n}Q=0$.\\
Now we re-index $F(z)$, obtaining

$$F'(\sqrt{z})=\sum_{j=-n}^{j=n}\tilde{C}_{2j}\sqrt{z}^{2j}.$$ 
 Crucially, in variable $\sqrt{z}$, $F'\in \xi_{2n}$. $F(z)=F'(\sqrt{z})$ on the unit circle, so this just has the effect of 'padding' $F$ with a lot of zero entries. We add superscript $\tilde{C}^{(2n)}_{2j}$ to denote that this is a coefficient in the $2n$-degree polynomial $F^{(2n)}(\sqrt{z})$, and identify $P_{2n}, F^{(2n-1)}$ from $\tilde{C}^{(2n)}_{2n}, (\tilde{C}^{(2n)})^{\dag}_{2n}$. 
 
Define $E_P=e^{i\theta/2}P+e^{-i\theta/2}Q$, where $\sqrt{z}=e^{in\theta/2}$ on the unit circle, and consider the expansion of $F(e^{i\theta/2})E_k(e^{i\theta/2})$,

\begin{align}
    F(e^{i\theta/2})E_P(e^{i\theta/2})&=e^{-i(n-1)\theta/2}\left(\tilde{C}_{2n-2}^{(2n)}P+\tilde{C}_{-2n+4}^{(2n)}Q\right)\nonumber\\
    &+...+e^{i(n-1)\theta/2}\left(\tilde{C}_{2n-4}^{(2n)}P+\tilde{C}_{2n}^{(2n)}Q\right)
\end{align} 
The leading terms are both $0$, and the definite parity of $\xi_n$ guarantees that $F(e^{i\theta/2})\xi_k(e^{i\theta/2})$ has definite parity $(n-1) mod2$. Thus $F(e^{i\theta/2})E_P(e^{i\theta/2})\in\xi_{n-1}$\footnote{If we instead take $E_P=e^{i\phi/2}L+e^{-i\phi/2}K$, then the leading terms $\tilde{C}_{2n}^{(2n)}, \tilde{C}_{-2n}^{(2n)}$ will add, rather than cancel, and we would get something in $\xi_{n+1}$}. Then using $F^{(2n-1)}$, we find $P_{2n-1}$ and $F^{(2n-2)}$ though precisely the same argument. Inductive steps for $2n, 2n-1..1$ gives us a unique set of projectors $\{P_1, P_2,...P_{2n}\}$ which defines $E_{P_1}(e^{in\theta/2})E_{P_2}(e^{in\theta/2})...\propto F'(e^{in\theta/2})$, and the remaining $F^{1}\in SU(2)$ defines $E_0$. 

\subsubsection{Step 4: Measurement}
We have found a decomposition
\begin{equation}
    F'(e^{in\theta/2})=E_0E_{P_1}(e^{in\theta/2})E_{P_2}(e^{in\theta/2})...E_{P_{2n}}(e^{in\theta/2})
\end{equation}
 that can prepare 
$$\begin{bmatrix}
    \mathcal{A}(z)+i\mathcal{D}(z) & -\mathcal{C}(z)+i\mathcal{B}(z)\\
    \mathcal{C}(z)+i\mathcal{B}(z) & \mathcal{A}(z)-i\mathcal{D}(z)
\end{bmatrix}.$$
For any $z\in U(1)$, measuring in the $\sigma_x$ basis and post-selecting $\ket{+}$ extracts
\begin{align}
    \bra{+}E_0E_{P_1}(e^{in\theta/2})E_{P_2}(e^{in\theta/2})...E_{P_{2n}}(e^{in\theta/2})\ket{+}
    =\mathcal{A}(z)+i\mathcal{B}(z).
\end{align}

\subsection{Error Analysis}\label{sec:QSP_haah_error_analysis}
In the previous section, we restated a derivation of the idealized Laurent-QSP circuit following \cite{Haah2019product}. However in practice, we have to manage two sources of error: error in the approximation and the limited precision in the solution to the Fejer problem. Error in the coefficients arises from two sources. First, the approximation for our target function $f(x)$ introduces some error. Secondly, there are QSP-relevant polynomials for which we cannot accurately represent the polynomial coefficients with floating point arithmetic. The rectangle function approximation introduced in \cite{low_quantum_2017} and discussed for QSP-processing in \cite{skelton2024harmlessmethodsqspprocessinglaurent} is paradigmatic of this problem.\\

\subsubsection{Coefficient and Functional Error Bounds}
Sometimes it is convenient to think about error attached to each coefficient of a polynomial object, but generally we would prefer to characterize error at a functional level with the $l^{\infty}$ norm. In the case of the polynomials we consider, the two can be directly related.
\begin{lemma}[Relating coefficient and functional error on $U(1)$]\label{lemma:fcn_to_coeff_error}
    Consider two $2\pi$-periodic Laurent polynomials of degree $n$, $g(z)=\sum_{-n}^ng_j z^j, h(z)=\sum_{j=-n}^nh_jz^j$, whose difference on $U(1)$ is bounded by $\epsilon\in\mathbb{R}_+$. In other words,
    $$\linfnorm{g-h}\leq \epsilon.$$
    Then for all $j\in \{-n, -n+1,...n-1, n\}$, we have that 
    $|g_j-h_j|\leq \epsilon$.
\end{lemma}
\begin{proof}
We interpret $g(z)-h(z)$ as a Fourier series in variable $\theta$, so that
\begin{align}\label{eq:fourier_coeff_bound_calc1}
   \left|g_k-h_k\right| &= \left|\frac{1}{2\pi}\int_{-\pi}^{\pi} \left(g(\theta)-h(\theta)\right)e^{-i\theta k}d\theta\right|\\
     &\leq \frac{1}{2\pi}\int_{-\pi}^{\pi} \left|\left(g(\theta)-h(\theta)\right)e^{-i\theta k}\right|d\theta\\
     &\leq \frac{1}{2\pi}\left(\int_{-\pi}^{\pi} \left|\left(g(\theta)-h(\theta)\right)\right|^2d\theta\right)^{1/2}\nonumber\\
     &\cdot \left(\int_{-\pi}^{\pi} \left|e^{-i\theta k}\right|^2d\theta\right)^{1/2}\label{eq:fourier_coeff_bound_calc2}\\
      &\leq \frac{1}{2\pi}\sqrt{2\pi}\epsilon\cdot \sqrt{2\pi}\leq \epsilon
\end{align}
\cref{eq:fourier_coeff_bound_calc2} follows from a simple version of  H{\"o}lder's inequality, since $g(t)-h(t)$ is square integrable on $U(1)$. The $l^{\infty}$ norms are bounded so trivially the $l_2$ norms must be as well. Finally note that since Schlatten norms also satisfy H{\"o}lder's inequality, \cref{eq:fourier_coeff_bound_calc1}-\cref{eq:fourier_coeff_bound_calc2} still holds when we replace $|\cdot|$ with the matrix operator norm and $\linfnorm{\cdot}$ with the operator norm over all $U(1)$ and work with matrix-valued coefficients.
\end{proof}

We formalize the direct connection between numerical polynomial representation and target function in \cref{lemma:targfcn_to_polylists}.
\begin{remark}[Arithmetic polynomial approximation of a target function]\label{lemma:targfcn_to_polylists}
    Let $f$ be a function which is $\epsilon_f$ close to $\mathcal{A}(z)+i\mathcal{B}(z)$, where $\mathcal{A, B}$ satisfy the conditions in \cref{lemma:QSP_Haah}. Assume we can access $\epsilon_{coeff}/2(2n+1)$ approximations to each coefficient of $\mathcal{A,B}$.  We further assume that the coefficient approximations build polynomial approximations $\tilde{\mathcal{A}}(z), \tilde{\mathcal{B}}(z)$ which preserve the real-on-circle, pure properties. Then, 
    \begin{equation}
        \left|\left|f-\left(\tilde{\mathcal{A}}+i\tilde{\mathcal{B}}\right)\right|\right|_{\infty}\leq \epsilon_f+\epsilon_{coeff}=\epsilon_{approx}.
    \end{equation}
\end{remark}
 \begin{proof}
We begin by assuming a set of Laurent polynomials $\mathcal{A}(z), \mathcal{B}(z)$, which approximate the desired function $\mathcal{F}(z)$ to some $\epsilon_f$. We allow the encoded representations of the coefficient lists to introduce additional error - assume the coefficient lists prepared numerically obey 
\begin{align}
    \left|{\mathcal{A}}_i-\tilde{\mathcal{A}}_i\right|\leq \frac{\epsilon_{coeff}}{2(2n+1)},
    \quad \left|{\mathcal{B}}_i-\tilde{\mathcal{B}}_i\right|\leq \frac{\epsilon_{coeff}}{2(2n+1)}
\end{align}
Crucially,  these coefficient errors do not change the properties of $\mathcal{A, B}$ required above. They are still (anti-)reciprocal, bounded, and have degrees at most $n$. $\tilde{\mathcal{A}}(z), \tilde{\mathcal{B}}(z)$ will remain close to ${\mathcal{A}}(z), {\mathcal{B}}(z)$, specifically
\begin{align}
    \linfnorm{{\mathcal{A}}-\tilde{\mathcal{A}}}&\leq \sum_i \left|{\mathcal{A}}_i-\tilde{\mathcal{A}}_i\right|\leq \frac{(2n+1)\epsilon_{coeff}}{{2(2n+1)}},\\
    \linfnorm{{\mathcal{B}}-\tilde{\mathcal{B}}}&\leq \frac{(2n+1)\epsilon_{coeff}}{{2(2n+1)}}.
\end{align}
Then, simple triangle-bound arguments give
$$\left|\left|f-\left(\tilde{\mathcal{A}}+i\tilde{\mathcal{B}}\right)\right|\right|_{\infty}\leq \epsilon_f+\epsilon_{coeff}.$$
 \end{proof}

We have explicitly separated these errors because $\epsilon_f$ is easily set by the user, whereas $\epsilon_{coeff}$ must be numerically computed for each instance or carefully bound using knowledge of the particular coefficient lists. However in practice, these can often be conflated into $\epsilon_{approx}$, and we will do so for the proof of \cref{thrm:laurent_qsp_with_processing}.

\subsubsection{Decomposition Step Error}
For the decomposition step, we do not directly care about the distance from the target function, and so we disregard $\epsilon_{coeff}$. Instead, we want to know how the error from the Fejer approximation $\mathcal{F}$ built from $\tilde{\gamma}$ instead of ideal ${\gamma}$ in \cref{sec:haah_proof_step3} will propagate to the QSP circuit. We will need a lemma to assist,
\begin{lemma}\label{lemma:helper_for_iteration}
    Assume a $2x2$ matrix valued function $\tilde{F}=\sum_{j=-n}^n \tilde{C}_jz^j\in\xi_n$ and some $\epsilon\in [0, \frac{1}{2})$, and then define $P=\frac{\tilde{C}_{-n}^{\dag}\tilde{C}_{-n}}{\text{Tr}\left(\tilde{C}_{-n}^{\dag}\tilde{C}_{-n}\right)}$. If the following conditions hold,
    \begin{enumerate}
        \item  $\opnorm{\tilde{F}(z)\left(\tilde{F}(z)\right)^{\dag}-F(z)\left(F(z)\right)^{\dag}}\leq \epsilon$ for all $z\in U(1)$, where $F(z)$ is also a degree n, matrix valued Laurent polynomial and is unitary for all $z\in U(1)$.  We say that $\tilde{F}=\sum_{j=-n}^n \tilde{C}_jz^j$ is "$\epsilon$-close to unitary". 
        \item For leading coefficient $\tilde{C}_{-n}$, $\opnorm{\tilde{C}_{-n}^{\dag}}\geq  c$  for some constant $c> 0$
        \item there exists $f>0$ such that $\opnorm{\Tilde{F}}\leq f$  for all $z\in U(1)$
    \end{enumerate}
  We have the following,
    \begin{enumerate}
        \item $\opnorm{P-\tilde{P}}\geq \frac{2\epsilon}{c}$ for some projector $\tilde{P}$,
        \item $\tilde{F}^{n-1}(z)=\tilde{F}(z)E_{\tilde{P}}(z)$ is $\epsilon$ close to unitary and $\tilde{F}^{n-1}(z)\in \xi_{n-1}$
        \item  For leading coefficient $\tilde{C}_{-n+1}^{(n-1)}$, $\opnorm{(\tilde{C}_{n-1}^{(n-1)})^{\dag}}\geq c$
        \item $\tilde{F}^{n-1}(z)$ is $\frac{4f\epsilon}{c}$-close to $\tilde{F}(z)E_{{P}}(z)$ and $\opnorm{\tilde{F}^{n-1}(z)}\leq f$ for all $z\in U(1)$ 
    \end{enumerate}
\end{lemma}
\begin{proof}
 $\Tilde{F}(z)\left(\Tilde{F}(z)\right)^{\dag}$ is also a Laurent polynomial, one with coefficients $\Tilde{L}_j$. From our first assumption, ${F}(z)\left({F}(z)\right)^{\dag}=I$ is a polynomial where $L_{-2n,}=L_{2n}=0$. Then using \cref{lemma:fcn_to_coeff_error},
$\opnorm{L_{j}-\Tilde{L}_j}\leq \epsilon$
and so
$$\opnorm{\Tilde{L}_{n}}, \opnorm{\Tilde{L}_{-n}}\leq \epsilon.$$
Denote the coefficients of $\Tilde{F}(z)$ by $\tilde{C}_{j}$. The following calculation shows that the action of $P=\frac{\tilde{C}_{-n}^{\dag}\Tilde{C}_{-n}}{\text{Tr}\left({\Tilde{C}_{-n}^{\dag}\Tilde{C}_{-n}}\right)}$ on $\Tilde{C}_{n}^{\dag}$ is close to the action of a projector $\Tilde{P}$ such that $\Tilde{P}\Tilde{C}_{n}^{\dag}=0$.
\begin{align}
 &\opnorm{\Tilde{P}\Tilde{C}_{n}^{\dag}-\frac{\Tilde{C}_{-n}^{\dag}\Tilde{C}_{-n}\Tilde{C}_{n}^{\dag}}{\text{Tr}\left({\Tilde{C}_{-n}^{\dag}\Tilde{C}_{-n}}\right)}}\\
 &=\opnorm{\Tilde{P}\Tilde{C}_{n}^{\dag}-\frac{\Tilde{C}_{-n}^{\dag}}{\text{Tr}\left({\Tilde{C}_{-n}^{\dag}\Tilde{C}_{-n}}\right)}\Tilde{L}_{n}}\\
    &\leq \frac{\opnorm{\Tilde{C}_{-n}^{\dag}}}{\text{Tr}\left({\Tilde{C}_{-n}^{\dag}\Tilde{C}_{-n}}\right)}\epsilon\leq \frac{\epsilon}{\opnorm{\Tilde{C}_{-n}^{\dag}}}=\frac{\epsilon}{c}\label{eq:approx_proj_bound}
\end{align}
We have two cases - in the first, $P$ is diagonalizable, so $P=e_0P_0+e_1P_1$ where $P_0, P_1$ are projectors built from each eigenvector of $P$. Because we have shown that $P$ is $\epsilon/c$-close to a projector in \cref{eq:approx_proj_bound}, and from the uniqueness of the eigenvector decomposition, we must have $|e_0|\leq \epsilon/c, |e_1-1| \leq \epsilon/c,$ or $|e_1|\leq \epsilon/c, |e_0-1| \epsilon/c$. Then we will define projector $\tilde{P}=P_1$ or  $\tilde{P}=P_0$ respectively and $\opnorm{P-\Tilde{P}}\leq \frac{\epsilon}{c}$. If $P$ is non-diagnolizable, then we show that $PP^{\dag}$ is close to the action of $\tilde{P}\tilde{P}^{\dag}=\tilde{P}$
\begin{align}
    \opnorm{PP^{\dag}\Tilde{C}_{n}^{\dag}-\tilde{P}\tilde{P}^{\dag}\Tilde{C}_{n}^{\dag}}&\leq \opnorm{P} \opnorm{P^{\dag}\Tilde{C}_{n}^{\dag}-\tilde{P}^{\dag}\Tilde{C}_{n}^{\dag}}\nonumber\\
    &+ \opnorm{P\Tilde{C}_{n}^{\dag}-\tilde{P}\Tilde{C}_{n}^{\dag}} \opnorm{\tilde{P}^{\dag}}\\
    &\leq \frac{2\epsilon}{c}.
\end{align}
In this case, $PP^{\dag}$ is Hermitian and can be used to define projector $\tilde{P}$. 

Because $E_{\tilde{P}}$ is strictly unitary, $\Tilde{F}(z)E_{\Tilde{P}}$ remains $\epsilon$-close to unitary, and we define $\Tilde{F}^{n-1}(z)=\Tilde{F}(z)E_{\Tilde{P}}$.  Note that $\opnorm{\Tilde{E}_{\Tilde{P}}}\leq 1$, easily derived from the definition of $\Tilde{P}$. 
The (possible) coefficients of $\tilde{F}^{n-1}(z)$ are
\begin{align}
    \tilde{C}_{n+1}^{(n-1)}&=\tilde{C}_{n}^{(n)}\tilde{P}=0, \quad \tilde{C}_{-n+1}^{(n-1)}=\tilde{C}_{-n+2}^{(n)}(1-\tilde{P})+\tilde{C}_{-n}^{(n)}\tilde{P},\nonumber\\
    \tilde{C}_{-n+3}^{(n-1)}&=\tilde{C}_{-n+4}^{n}(I-\tilde{P})+\tilde{C}_{-n+2}\tilde{P},...\nonumber\\
    \tilde{C}_{n-1}^{(n-1)}&=\tilde{C}_{n}^{(n)}(1-\tilde{P})+\tilde{C}_{n-2}^{(n)}\tilde{P},\quad \tilde{C}_{n+1}^{(n-1)}=\tilde{C}_{n}^{(n)}\tilde{P}=0
\end{align}
From the definitions of the projector $\tilde{P}$,$\tilde{C}_{-n-1}^{(n-1)}=e_1P_1\leq \epsilon P_1$, $\tilde{C}_{n+1}^{(n-1)}=0$, which is adequate for our purposes to say $\tilde{F}^{n-1}(z)\in \xi_{n-1}$.

The last-leading coefficient is $\tilde{C}_{-n+1}^{(n-1)}$, and its operator norm bound comes from observing that $\tilde{F}^{(n-1)}$ is the product of some series $E_p$ as defined in \cref{sec:haah_proof_step3}, and the leading coefficient has the form $\tilde{C}_{n-1}^{(n-1)}=E_0P_1P_2...\Tilde{P}_{n-1}$, or under conjugation, $(\tilde{C}_{n-1}^{(n-1)})^{\dag}=\Tilde{P}_{n-1}^{\dag}...P_1^{\dag}E_0^{\dag}$. Although at this point we cannot assume this is a series of exact projectors, we can use its existence and the sub-multiplicity of the operator norm to argue that $\opnorm{(\tilde{C}_{n-1}^{(n-1)})^{\dag}}\geq\opnorm{(\tilde{C}_{n}^{(n)})^{\dag}}\geq c$. 
Finally, after the transformation,
\begin{align}
    \opnorm{\Tilde{F}(z)E_{{P}}-\Tilde{F}(z)E_{\Tilde{P}}}&\leq \opnorm{\Tilde{F}(z)}\opnorm{E_{{P}}-E_{\tilde{P}}}\\
    &\leq 2\opnorm{\Tilde{F}(z)}\opnorm{P-\tilde{P}}\\
    &\leq 4f\frac{\epsilon}{c}
\end{align}
and $\opnorm{\tilde{F}^{(n-1)}}\leq \opnorm{\tilde{F}}\opnorm{E_{\Tilde{P}}}\leq f$.
\end{proof}

\begin{theorem}[Decomposition Step proof]\label{lemma:decomp_error}
    Given $\mathcal{\Tilde{A}}, \mathcal{\Tilde{B}}$ such that the leading coefficients satisfy  $|\mathcal{\Tilde{A}}_{n}|, |\mathcal{\Tilde{B}}_{n}|, |\mathcal{\Tilde{A}}_{-n}|, |\mathcal{\Tilde{B}}_{-n}|\geq \frac{c}{2}$ for some $c>0$  
    Assume access to Fejer problem solution $ \Tilde{\gamma}$ such that $\linfnorm{\gamma-\Tilde{\gamma}}\leq \epsilon_{fejer}$ for $\epsilon_{fejer}\in [0, \frac{c}{2})$ we have that
    $$\linfnorm{1-\Tilde{\mathcal{A}}^2(z)-\Tilde{\mathcal{B}}^2(z)-\tilde{\gamma}(z)\Tilde{\gamma}(z^{-1})}\leq 4\linfnorm{\Tilde{\gamma}}\epsilon_{fejer},$$ 
    Furthermore, there exists a set of projectors $\{\Tilde{P}\}$ defining $E_{\Tilde{P}}$'s such that
    $$\opnorm{\bra{+}\prod E_{\Tilde{P}}(z)\ket{+}-\mathcal{\Tilde{A}}(z)+i\mathcal{\Tilde{B}}(z)}\leq c_n\epsilon_{fejer}$$
    for all $z\in U(1)$, where ${c}_n=\frac{4}{c}(2n+1)2^{2n}$.
\end{theorem}

\begin{proof}
    Because $1-\mathcal{\Tilde{A}}^2- \mathcal{\Tilde{B}}^2$ fits the conditions required in \cref{def:fejer_fact}, there exists a solution $\gamma$ such that $1-\mathcal{\Tilde{A}}^2(z)- \mathcal{\Tilde{B}}^2(z)=\gamma(z)\gamma(z^{-1})$. Then we can idealize the following
    \begin{align}
        \mathcal{{C}}(z)=\frac{\gamma(z)+\gamma(z^{-1})}{2}\\
        \mathcal{D}(z)=\frac{\gamma(z)-\gamma(z^{-1})}{2i}.
    \end{align}
    We assume access to a solution of the Fejer problem $\tilde{\gamma}$ as in \cref{sec:wilson}. 
    Then, again using simple triangle bounds and $\linfnorm{{\gamma}}\leq (1+\varepsilon_{fejer})\linfnorm{\tilde{\gamma}}\leq 2\linfnorm{\tilde{\gamma}}$,
$$\linfnorm{1-\Tilde{\mathcal{A}}^2(z)-\Tilde{\mathcal{B}}^2(z)-\tilde{\gamma}(z)\Tilde{\gamma}(z^{-1})}\leq 4\varepsilon_{fejer}\linfnorm{\gamma}.$$ 
    
    From \cref{sec:haah_proof_step2}, we are guaranteed that $ F(z)=\mathcal{\Tilde{A}}(z)I+i\mathcal{\Tilde{B}}(z)\sigma_x+i\mathcal{C}(z)\sigma_y+i{\mathcal{D}}(z)\sigma_z$ obeys $F(z)\left(F(z)\right)^{\dag}=I$. $F(z)\left(F(z)\right)^{\dag}$ is also a Laurent polynomial with coefficients $L_{j}$, and we must have $L_{-2n}, L_{2n}=0$.
    However we actually instantiate polynomials defined from the Wilson method solution,
    \begin{align}
         \mathcal{\Tilde{C}}(z)=\frac{\Tilde{\gamma}(z)+\Tilde{\gamma}(z^{-1})}{2}\\
        \mathcal{\Tilde{D}}(z)=\frac{\Tilde{\gamma}(z)-\Tilde{\gamma}(z^{-1})}{2i},
    \end{align}
    from which we build 
    $$\Tilde{F}(z)=\mathcal{\Tilde{A}}(z)I+i\mathcal{\Tilde{B}}(z)\sigma_x+i\mathcal{\Tilde{C}}(z)\sigma_y+i{\mathcal{\Tilde{D}}}(z)\sigma_z.$$
We now want to determine whether $\tilde{F}$ is close to unitary in the operator norm for all $z\in U(1)$, as characterized by the following, 
\begin{align}
    \opnorm{I-\Tilde{F}(z)\left(\Tilde{F}(z)\right)^{\dag}}\leq \opnorm{\Tilde{F}^{\dag}}\cdot \opnorm{F-\Tilde{F}}+\opnorm{F}\cdot \opnorm{\left(F\right)^{\dag}-\left(\Tilde{F}\right)^{\dag}}.
\end{align}
We have that $\opnorm{\mathcal{C}\sigma_Y-\mathcal{\tilde{C}}\sigma_Y}\leq \linfnorm{\mathcal{C}-\mathcal{\Tilde{C}}}\cdot \opnorm{\sigma_Y}=\epsilon_{fejer}$ and a similar expression holds for $\mathcal{D}, \mathcal{\tilde{D}}$. Then $\opnorm{F-\Tilde{F}}\leq 2\epsilon_{fejer}$ for all $z\in U(1)$, and because $\mathcal{C, D}$ are real-on-circle, the same bound holds for $\opnorm{F^{\dag}-\Tilde{F}^{\dag}}$. We can also ensure from the properties of $\mathcal{\tilde{A},\tilde{B}, C, D}$ that $\opnorm{F}\leq 1$ and then $\opnorm{\Tilde{F}}\leq 1+2\epsilon_{fejer}$ for all $z\in U(1)$. For all $z\in U(1)$,
\begin{equation}
    \opnorm{I-\Tilde{F}(z)\left(\Tilde{F}(z)\right)^{\dag}}\leq 4\epsilon_{fejer}+4\epsilon_{fejer}^2\leq 8\epsilon_{fejer}.
\end{equation}
    
Now we are ready to determine how this error carries through the decomposition. First we need to lowerbound $\opnorm{(\Tilde{C}_{-2n})^{\dag}}$, which naturally emerges from our condition on the coefficient lists
\begin{align}
\opnorm{\Tilde{C}_{-2n}^{\dag}}\geq  \left|\mathcal{\tilde{A}}^{\dag}_{-2n}\right|+\left|\mathcal{\tilde{B}}_{-2n}^{\dag}\right|\geq c
\end{align}
From \cref{lemma:helper_for_iteration} ,
$$P=\frac{\left(\tilde{C}_{-2n+1}^{(2n-1)}\right)^{\dag}\tilde{C}_{-2n+1}^{(2n-1)}}{\text{Tr}\left(\left(\tilde{C}_{-2n+1}^{(2n-1)}\right)^{\dag}\tilde{C}_{-2n+1}^{(2n-1)}\right)}$$ can be used to identify projector  $\Tilde{P}$, preparing $\Tilde{F}^{2n-1}(z)$ which is $f\epsilon$-close to the idealized ${F}^{2n-1}(z)$ defined in \cref{sec:haah_proof_step3}.

 Now we are free to proceed as in Step 3, computing projectors $\Tilde{P}_{2n}, \Tilde{P}_{2n-1},...\Tilde{P}_1, \Tilde{E}_0$ within bounded error. At each step, $\tilde{C}_{-j}^{(j)}$ remains bounded.

We must verify that this sequence remains close to $\Tilde{F}(z)=\prod E_P(z)$ where each $E_P$ is a term in the exact (ie not exactly unitary) expansion. Then using straightforward identities, we must have the following for all $z\in U(1)$
\begin{align}
\opnorm{\prod_{j=0}^{2n}\Tilde{E}_{\Tilde{P}}-\tilde{F}}&=\opnorm{\prod_{j=0}^{2n}\Tilde{E}_{\Tilde{P}}-\prod_{j=0}^{2n}{E}_{P}}\\
&\leq \opnorm{\Tilde{E}_0-E_0}+\sum_{j=1}^{2n}\left(\prod_{k=0}^j\opnorm{E_k}\right)\cdot\opnorm{\Tilde{E}_j-E_j}\\
    &\leq \sum_{j=0}^{2n}\prod_{k=0}^j\left(1+2\opnorm{P_k-\tilde{P}_k}\right)\cdot 2\opnorm{P_j-\tilde{P}_j}\\
    &\leq \frac{4\epsilon_{fejer}}{c}\sum_{j=0}^{2n}\prod_{k=0}^j\left(1+\frac{4\epsilon_{fejer}}{c}\right)\\
    &\leq \frac{4\epsilon_{fejer}}{c}\sum_{j=0}^{2n}\left(1+\frac{4\epsilon_{fejer}}{c}\right)^j\\
    &\leq \frac{4(2n+1)\epsilon_{fejer}}{c}\left(1+\frac{4\epsilon_{fejer}}{c}\right)^{2n}\\
    &\leq \frac{4(2n+1)}{c}\epsilon_{fejer}\sum_{k=0}^{2n}\binom{2n}{k}\left(\frac{4\epsilon_{fejer}}{c}\right)^{k}\\
    &\leq \frac{4(2n+1)\epsilon_{fejer}}{c}\sum_{k=0}^{2n}\binom{2n}{k}\\
\end{align}

Then finally,
\begin{align}
    \opnorm{\prod_{j=1}^{2n}E_{\tilde{P}}(\sqrt{z})-\Tilde{F}({z})}&\leq \frac{4(2n+1)}{c}2^{2n}\epsilon_{fejer}
\end{align}

\end{proof}
\subsection{Proof of Theorem 2}
We combine the results of the previous two lemmas to determine the full error in our QSP procedure,
\LQSPTHRM*
\begin{proof}
The error on each coefficient means that that the error in $\mathcal{\Tilde{A}, \tilde{B}}$ is at most $\frac{\epsilon}{2}$. Then using an initial guess determined by \cref{lemma:wilson-convergence}, we can identify a $\tilde{\gamma}$ solution using \cref{lemma:wilson} to precision $\epsilon_{fejer}$. Combining the results of \cref{lemma:targfcn_to_polylists} and \cref{lemma:decomp_error},
\begin{align}
     \linfnorm{1-f(z)f^{*}(z)-\bra{+}\prod_{j=1}^{2n}E_{\Tilde{P}}(\sqrt{z})\ket{+}}
     &\leq  \linfnorm{1-f(z)f^{*}(z)-\mathcal{\tilde{A}}-i\mathcal{\Tilde{B}}}\nonumber\\
     &+ \linfnorm{\mathcal{\tilde{A}}+i\mathcal{\Tilde{B}}-\bra{+}\prod_{j=1}^{2n}E_p(\sqrt{z})\ket{+}}\\
     &\leq \frac{\epsilon}{2}+c_n\epsilon_{fejer}\leq \epsilon.
\end{align}
\end{proof}

\section{QSP Circuit Basics}
\subsection{Controlled QSP Circuits}
\label{sec:LCU_QSVT}
The argument given below is a synthesis of existing literature. We review how the controlled versions of G-QSP can be run in parallel\footnote{We chose to show this result in G-QSP because then the result holds for all QSP conventions. However, since G-QSP already implements any $P\in mathbb{C}$ such that $\linfnorm{P}\leq 1$, this result is not useful by itself.}, leading to a theorem similar (and extendable to) the QSVT case given in \cite{gilyen_quantum_2019}. The template for adding a series of functions $f_i$ fitting the requirements of \cref{lemma:QSP_gilyen}, 'addition of block-encodings', originates in \cite{gilyen_quantum_2019}. A similar technique for QSP and Hermitian block-encodings is given in \cite{dong_efficient_2021}. 
Recall that the circuit is 
$$U_{\Vec{\theta}, \Vec{\phi}}=\left(\prod_{j=1}^dR(\theta_j, \phi_j, 0)A\right)R(\theta_0, \phi_0, \lambda)=\begin{bmatrix}
            {P}(U) & \cdot\\
            \cdot & \cdot\\
        \end{bmatrix}.$$
In this section we will be imprecise with notation: $R_i$ is any signal processing operator,  $C_kU$ is the controlled operation $\left(I-\ket{k}\bra{k}\right)\otimes I+\ket{k}\bra{k}\otimes U$ and $CU$ is the signal operator $C_1U$ controlled on the QSP-ancilla qubit, $U_{R_i}$ is one QSP step, and $U_{P}$ is the QSP circuit preparing $\sum_{\theta}P(e^{i\theta})\ket{\theta}\bra{\theta}$. We will use a G-QSP circuit that alternates $CU, C^{\dag}$ operations. It does not matter whether the oracle is $CU, CU^{\dag}$ or $CU+CU^{\dag}$ \\

Each step of the circuit has the form \cref{fig:GQSP_circuit_step}, and one can easily verify that \cref{fig:CGQET_circuit_2steps} gives two subsequent QSP steps, controlled on an additional ancilla qubit. 

So then we build successive controlled steps by controlling \textit{the signal processing operators only}, as in \cref{fig:CGQET_circuit_2steps}. If the circuit is even in length, then we are done. If the sequence is odd length, then we will need to implement $C_{11}U.$ The full controlled odd circuit is \cref{fig:odd_GQSP}. 

\begin{figure}[ht]
\begin{subfigure}[b]{0.3\textwidth}
    \centering
    \begin{quantikz}\
        \qw & \gate[2]{U_{R_i}} & \qw\\
         \qwbundle[alternate]{} & \qwbundle[alternate]{} & \qwbundle[alternate]{}
    \end{quantikz}=
    \begin{quantikz}
\qw & \ctrl{1} & \gate{R_{i}} & \qw\\
\qwbundle[alternate]{} & \gate{U} \qwbundle[alternate]{} & \qwbundle[alternate]{} & \qwbundle[alternate]{}
\end{quantikz}
    \caption{step of a G-QSP circuit}
    \label{fig:GQSP_circuit_step}
\end{subfigure}
\begin{subfigure}[b]{0.3\textwidth}
\centering
    \begin{quantikz}
        \qw & \ctrl{1} & \qw\\
        \qw & \gate[2]{U_{\{R_i, R_{i+1}\}}} & \qw\\
         \qwbundle[alternate]{} &  \qwbundle[alternate]{} & \qwbundle[alternate]{}
    \end{quantikz}=
    \begin{quantikz}
    \qw & \qw & \ctrl{1} & \qw  & \ctrl{1} & \qw \\
    \qw &  \ctrl{1} & \gate{R_{i}} & \ctrl{1} & \gate{R_{i+1}} &\qw\\
    \qwbundle[alternate]{} & \gate{U} \qwbundle[alternate]{} & \qwbundle[alternate]{}  & \gate{U} \qwbundle[alternate]{} &  \qwbundle[alternate]{} &  \qwbundle[alternate]{}
\end{quantikz}
    \caption{Controlled successive Steps of a G-QET circuit}
    \label{fig:CGQET_circuit_2steps}
\end{subfigure}
\end{figure}

The other QSP cases are nearly identical - the standard circuit \cref{fig:QET_basic_circuit} can be made structurally identical to QSP by redefining $V_{i}^{\dag}V_{i+1}$ pairs into $R_i$ (Laurent-QSP) or restricting $R_i$ to be a specified Pauli rotation (angle-QSP). Then, our argument proceeds as before, although at this point, the set of expressible functions remains constrained. Similarly, the same circuit is usable for QSVT whenever $CU$ is replaced with $C\Pi(U_{BE})C\Pi$ for block-encoding $U_{BE}$ defined with the relevant singular values.

\subsection{Linear Combination of Unitaries for QET, QSVT}\label{sec:proof-LCU-circuits}
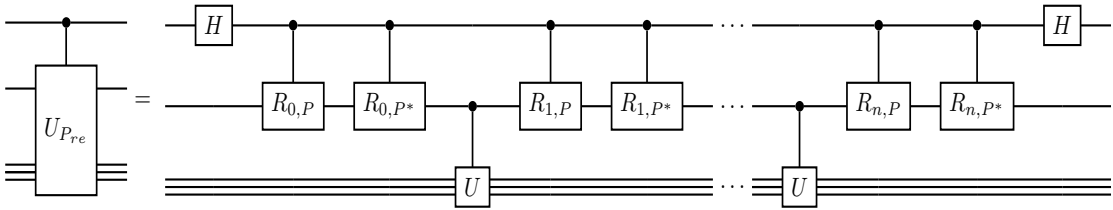
\begin{figure}[ht]
   \scalebox{0.8}[1]{
    \begin{quantikz}
        \qw & \ctrl{1} & \qw\\
        \qw & \gate[2]{U_{P_{re}}} & \qw\\
         \qwbundle[alternate]{} & \qwbundle[alternate]{} & \qwbundle[alternate]{}
    \end{quantikz}=
    \begin{quantikz}
    \qw & \gate{H} &\ctrl{1} & \ctrl{1} & \qw  & \ctrl{1} & \ctrl{1} &  \qw \ \ldots\ & \qw & \ctrl{1}  &  \ctrl{1} &  \gate{H} & \qw\\
\qw & \qw & \gate{R_{0, P}} & \gate{R_{0, P^*}} & \ctrl{1} & \gate{R_{1, P}} & \gate{R_{1, P^*}} &  \qw \ \ldots\  &  \ctrl{1} & \gate{R_{n, P}} & \gate{R_{n, P^*}} & \qw & \qw\\
\qwbundle[alternate]{} & \qwbundle[alternate]{} & \qwbundle[alternate]{} & \qwbundle[alternate]{}  & \gate{U} \qwbundle[alternate]{} &\qwbundle[alternate]{}& \qwbundle[alternate]{} & \qwbundle[alternate]{}  \ \ldots\  &  \gate{U} \qwbundle[alternate]{} &\qwbundle[alternate]{}  &\qwbundle[alternate]{} & \qwbundle[alternate]{} & \qwbundle[alternate]{}
\end{quantikz}
}
    \caption{Controlled QSP circuit preparing $P_{real}=\frac{P+P^{\dag}}{2}$ where we assume the circuit is even}
    \label{fig:GQSP_real_circuit}
\end{figure}

Linear combination of unitaries techniques are extremely compatible with QSP and QSVT; we can use the fact that most signal processor steps do not require extra control qubits to build up to $2^m$ different QSP circuits at once. The most general results are given in \cite{gilyen_quantum_2019}, and the case where the block-encoding is Hermitian is given in \cite{dong_efficient_2021}. One highly relevant case is when we want to prepare the real part of a polynomial. Then working with any QSP convention, we can prepare $P^{\dag}$ with another degree $n$ QSP sequence, and so $\frac{1}{2}\left(U_{P}+U_{P^{\dag}}\right)=U_{P_{real}}$ can be prepared with \cref{fig:GQSP_real_circuit}.
 
We now provide a proof of \LCUQSP*
\begin{proof}
    wlog assume that $l_e=\log_2 L_e, l_o=\log_2 L_o$ are both integers. With access to $l_e$ ancillary qubits in addition to the QSP ancillary qubit, we can embed $L_e$ even QSP circuits specified with signal processing sets $\{R\}_k$, $k\in{1, 2...L_e}$ as 
\begin{equation}\label{eq:qsp_LCU_even}
    \sum_{k=1}^{L_e}\ket{k}\bra{k}U_{\{R\}_k} =\prod_{j=1}^{n} \sum_k \ket{k}\bra{k}C_kR_{j, k}CU.
\end{equation}
\cref{eq:qsp_LCU_even} can be implemented with $n$ calls to $CU, CU^{\dag}$, and $n+1$ steps applying $L_e$ rotation gates, each of which is controlled on a particular $k$. We assume the cost of a $0$ controlled operation and a $1$ controlled operation are identical (or negligible compared to the cost of the generalized Toffoli gate).\\
If we want odd QSP circuits of length $n+1$ and have access to $l_o$ ancillary qubits, then we have
\begin{align}
    \sum_{k=1}^{L_o}\ket{k}\bra{k}U_{\{R\}_k} &=\left(\prod_{i=1}^{n} \sum_k \ket{k}\bra{k}C_kR_{j, k}CU\right)\nonumber\\
    &\cdot\sum_{k}\ket{k}\bra{k} C_kR_{n, k}CU
\end{align}
Moreover, if we want to combine $n$ and $n\pm 1$ length circuits, we can do so by slightly modifying the structure of the control gates above. Say we assign the odd functions index labels from $\{1...L_o\}$ and the even functions labels within $\{L_o+1,...L_e+L_o\}$. The first $n$ steps are just as in the two cases above,
$$\sum_{k=1}^{L_e+L_o}\ket{k}\bra{k}U_{\{R\}_k} =\prod_{j=1}^{n} \sum_k \ket{k}\bra{k}C_kR_{j, k}CU.$$

Now, we want to implement the last step for the odd functions without affecting the even functions. We can do this by controlling $U$ over the $l_o$ qubits covering the odd polynomials, as well as the QSP-qubit. This operation can be straightforwardly implemented with $L_o$ different $C_{l_o+1}$ operations. The rotation gates for $L_o$ odd functions can be straightforwardly implemented, still in the $l_o$ qubits. 
When we measure the $l$ ancillas in the $X$-basis, we arrive at
\begin{align}
    \quad &\bra{0^{l}}H^{\otimes l}\sum_{k=1}^{L_1+L_2}\ket{k}\bra{k}U_{\{R\}_k}H^{\otimes m}\ket{0^{\otimes l}}=\frac{1}{L_1+L_2}\sum_{\theta}f(e^{i\theta})\ket{\theta}\bra{\theta}
\end{align}
\end{proof}

The key point is that the query complexity of $U, U^{\dag}$ will scale with $n$, but the subnormalization, and hence the success probability, scales with $\frac{1}{L_1+L_2}$. Correcting the subnormalization on the circuit with quantum amplitude amplification would require a multiplicative factor on the query complexity roughly proportional to the number of polynomials in the sum \cite{gilyen_quantum_2019}. Assuming that $1, 2$-controlled rotations are less expensive than the signal operator, then the choice of QET, QSVT convention should not significantly alter the complexity of this routine.

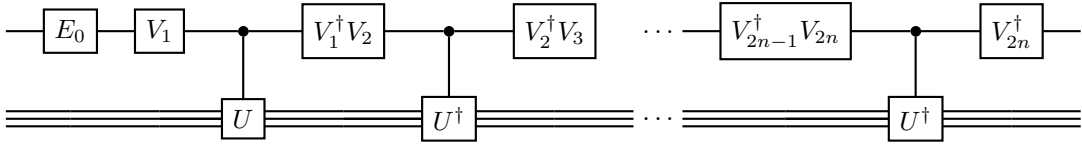
\begin{figure}[ht]
  \centering
       \begin{quantikz}
\qw & \gate{E_0} & \gate{V_1} & \ctrl{1} & \gate{V_1^{\dag}V_2} &  \ctrl{1}& \gate{V_2^{\dag}V_3} &\ \ldots\  & \gate{V^{\dag}_{2n-1}V_{2n}} & \ctrl{1}  & \gate{V_{2n}^{\dag}} & \qw\\
\qwbundle[alternate]{} & \qwbundle[alternate]{}  & \qwbundle[alternate]{}& \gate{U}\qwbundle[alternate]{} & \qwbundle[alternate]{} 
& \gate{U^{\dag}}\qwbundle[alternate]{} & \qwbundle[alternate]{} & \qwbundle[alternate]{} \ \ldots\ & \qwbundle[alternate]{}  &  \gate{U^{\dag}}\qwbundle[alternate]{}  &  \qwbundle[alternate]{}&  \qwbundle[alternate]{}
\end{quantikz}
 \caption{Basic QSP circuit for operator $U$,  preparing $\mathcal{A}+i\mathcal{B}: \{\lambda\}\rightarrow |z|<1$ given set of unitaries $\{V\}\in SU(2)^{\otimes n}$, $n=2k$.}
    \label{fig:QET_basic_circuit}
\end{figure}
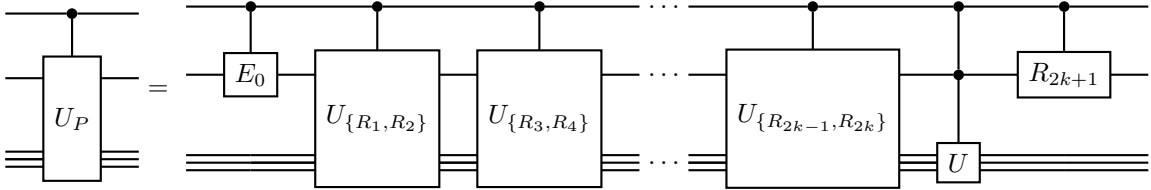
\begin{figure}[ht]
  \centering
   
        \scalebox{1}[1]{
    \begin{quantikz}
        \qw & \ctrl{1} & \qw\\
        \qw & \gate[2]{U_{P}} & \qw\\
         \qwbundle[alternate]{} & \qwbundle[alternate]{} & \qwbundle[alternate]{}
    \end{quantikz}=
    \begin{quantikz}
    \qw & \ctrl{1} & \ctrl{1}  & \ctrl{1} &  \qw \ \ldots\ & \ctrl{1}  &  \ctrl{1} & \ctrl{1} &\qw\\
\qw & \gate{E_0} & \gate[2]{U_{\{R_1, R_{2}\}}} & \gate[2]{U_{\{R_3, R_{4}\}}} &  \qw \ \ldots\  & \gate[2]{U_{\{R_{2k-1}, R_{2k}\}}} & \ctrl{1} &\gate{R_{2k+1}} & \qw\\
\qwbundle[alternate]{} & \qwbundle[alternate]{}  & \qwbundle[alternate]{} & \qwbundle[alternate]{} & \qwbundle[alternate]{}  \ \ldots\ & \qwbundle[alternate]{}  &  \gate{U} \qwbundle[alternate]{} &\qwbundle[alternate]{} &\qwbundle[alternate]{}
\end{quantikz}
}
    \caption{Controlled QSP circuit for $\{R\}\in SU(2)^{\otimes n}$, $n=2k+1$.}
    \label{fig:odd_GQSP}
\end{figure}

\end{document}